# Understanding the fill-factor limit of organic solar cells

Huotian Zhang[1*], Jun Yuan[2], Tong Wang[3], Nurlan Tokmoldin[4,5], Rokas Jasiūnas[6], Yiting Liu[7], Manasi Pranav[4], Yuxuan Li[1], Xiaolei Zhang[7], Vidmantas Gulbinas[6], Safa Shoaee[4,5], Yingping Zou[2], Veaceslav Coropceanu[8], Artem A. Bakulin[3], Dieter Neher[4*], Thomas Kirchartz[9,10], Feng Gao[1*]

1. Department of Physics, Chemistry and Biology (IFM), Linköping University, Linköping, Sweden

2. College of Chemistry and Chemical Engineering, Central South University, Changsha, China

3. Department of Chemistry and Centre for Processable Electronics, Imperial College London, London, United Kingdom

4. Optoelectronics of Disordered Semiconductors, Institute of Physics and Astronomy, University of Potsdam, Potsdam-Golm, Germany

5. Paul-Drude-Institut für Festkörperelektronik, Leibniz-Institut im Forschungsverbund Berlin, Berlin, Germany

6. Center for Physical Sciences and Technology, Vilnius, Lithuania

7. State Key Laboratory of Precision Spectroscopy, School of Physics and Electronic Sciences, East China Normal University, Shanghai, China

8. Department of Chemistry and Biochemistry, University of Arizona, Tucson, AZ, USA

9. IMD-3 Photovoltaics, Forschungszentrum Jülich, Jülich, Germany

10. Faculty of Engineering and CENIDE, University of Duisburg-Essen, Duisburg, Germany

* Corresponding Author: huotian.zhang@liu.se, neher@uni-potsdam.de, feng.gao@liu.se

# Abstract

Although the power conversion efficiencies of organic solar cells (OSCs) have surpassed 20%, they still lag behind commercial inorganic solar cells and emerging perovskite solar cells. To bridge this efficiency gap, improving the fill factor (*FF*) is critical, provided other photovoltaic parameters are not compromised. However, the fundamental understanding of the *FF* in OSCs remains incomplete. In this work, we systematically investigate a wide range of OSCs with the *FF* values spanning 0.27 to 0.80, and analyse the effect of free charge generation and recombination on the *FF* in OSCs. To explain our observations, we developed an analytical model that quantitatively correlates the applied electric field with the energetics of excited states in donor-acceptor blends. By combining device characterisation, spectroscopy, and theoretical modelling, we reveal that the Stark effect and the field-dependent charge transfer significantly impact the *FF* in state-of-the-art OSCs with low voltage losses. Our findings highlight that suppressing geminate decay by increasing exciton lifetime is a promising strategy for boosting the *FF* and achieving future efficiency gains in OSCs.

# Introduction

Organic solar cells (OSCs) have recently achieved impressive power conversion efficiencies (PCE) of > 20%[1,2]. These advances mainly benefit from the mitigated competition between open-circuit voltage ($V_{OC}$) and short-circuit current density ($J_{SC}$)[3–9] in OSCs based on non-fullerene acceptors (NFAs). Despite a thorough understanding and significant improvement in $V_{OC}$ and $J_{SC}$, the third parameter, the fill factor (*FF*), which is equally important for PCE, is not well understood for OSCs. A critical question is whether there is a trade-off between the $V_{OC}$ and *FF*.[10–12]

The *FF* in solar cells is complex, as it is influenced by multiple processes and quasi-equilibrium states that vary with voltage. Consequently, the relationship between the *FF* and $V_{OC}$ may vary depending on not only the type of solar cells, but even the observational methodologies. For example, some photovoltaic technologies shows an improved *FF* with a higher $V_{OC}$; this trend is typically attributed to decreased recombination losses (at the bulk heterojunction or at interfaces to contact layers), as observed in solution-processed perovskite solar cells[13–15]. However, contrasting behaviours have been observed in other high-efficiency solar cell technologies, where higher $V_{OC}$ values are accompanied by a reduced $FF^{16,17}$. For instance, in silicon solar cells, this phenomenon can result from the interplay between the contact recombination and contact resistance[16,18]. Therefore, understanding the *FF*-$V_{OC}$ relationship is a critical area of research across all solar cell technologies, including OSCs.

The anticorrelation between the *FF* and $V_{OC}$ in OSCs was not a primary concern until the recent development of NFAs. During the fullerene period, the voltage loss was the most significant limiting factor for the PCE of OSCs and drew the most attention[19,20]. With the latest advancements in NFAs, the voltage losses in OSCs have been significantly reduced. However, the improved $V_{OC}$ has also introduced concerns regarding the *FF*[10–12]. These observations raise the important question of whether there is an intrinsic limit to the *FF* of OSCs and how it correlates with voltage losses. A fundamental understanding of the *FF* limit in OSCs is needed for exploiting the full potential of organic semiconductors.

To understand the fundamental limits of the *FF* in OSCs, we systematically investigate its correlation with voltage losses across a diverse set of devices. For OSCs with $V_{OC}$ losses ranging from 0.5 eV to 1.1 eV and *FF* values spanning 0.27 to 0.80, we find that the *FF* of OSCs is not only dependent on the transport properties that determine the *FF* in conventional semiconductors but is also influenced by geminate recombination losses. Focusing on representative systems employing NFAs with minimal voltage losses, we probe the details of free charge generation and recombination under different voltages. These experimental data inform an analytical model that establishes field-dependent free charge generation as a pivotal factor for *FF* losses. We further connect this to the spectral evolution under bias, enabling detailed theoretical modelling of the electric field effect on photochemical dynamics in OSCs. Our findings converge on a critical conclusion: suppressing non-radiative exciton decay is the key to overcoming the *FF* limitations and unlocking the full potential of OSCs.

## Fill factor and open-circuit voltage

For solar cells with ideal extraction and without resistive losses, the analytical relationship between the *FF* and the $V_{OC}$ can be described by[21]

$$FF = \frac{v_{oc} - ln(v_{oc} + a)}{v_{oc} + 1} \quad (1)$$

and

$$v_{OC} = \frac{V_{OC}}{n_{id} k_B T} \quad (2)$$

where $v_{OC}$ is the normalized $V_{OC}$, $a$ is an empirical constant, $n_{id}$ is the ideality factor, $k_B$ is the Boltzmann constant and $T$ is the temperature. If $n_{id}$ is equal to 1, the solar cell is an ideal diode with a recombination order of 2, as indicated by the solid curve in **Figure 1a**. Note that high-quality Si may overcome this "classical" Auger limit of the *FF* (but not of the efficiency) [22], whereas the recombination in OSCs is typically considered to be bimolecular[23,24]. As predicted by this curve ($n_{id}$ = 1), *FF* increases with increasing $V_{OC}$, indicating that a solar cell with a higher $V_{OC}$ would also have a potentially higher *FF*. This is also a natural characteristic of diodes due to an exponential increase in their current with voltage.

For comparison, we summarise the $V_{OC}$ and *FF* data of diverse types of solar cells from the literature and present them in Figure 1a. As seen in Figure 1a, Si and perovskite solar cells are very close to the theoretical limit because of their efficient charge collection[25,26]. However, OSCs show dispersed data points, with widespread $V_{OC}$ ranging from 0.52 to 1.11 V and overall lower *FF* from 0.27 to 0.80. This is because they are composed of various organic molecules, resulting in diverse optical and transport properties. Therefore, it is difficult to describe the relationship between *FF* and $V_{OC}$ in OSCs.

To ease the analysis of the OSCs, we converted the $V_{OC}$ values into voltage losses, as shown in **Figure 1b**. In this context, voltage loss refers to the difference between the optical gap divided by the elementary charge ($E_{opt}/q$) and the $V_{OC}$. The method for extracting $E_{opt}$ is detailed in **Supplementary Figure 1** and the extracted values are listed in **Supplementary Table 1**. Upon the transformation of horizontal axis to voltage losses, the data dispersion due to variations in $E_{opt}$ is suppressed. Compared to $V_{OC}$, voltage losses provide a more accurate representation of the device performance based on varied materials. Although the data points remain scattered, we observe that, firstly, there is an upper limit of the *FF*, which increases with decreasing voltage losses, as indicated by the guide-to-the-eye yellow curve; and secondly, at small voltage losses, there is a sharp decrease in the *FF* values with decreasing voltage losses, a left boundary as demonstrated by the guide-to-the-eye green curve, indicating competition between $V_{OC}$ and *FF* in this region. We highlight the importance of understanding this left border, as it limits the *FF* (and hence the PCE) of low-voltage-loss OSCs. In other words, overcoming the limitations that currently cause this border would make it possible to further enhance the PCE of the state-of-the-art OSCs.

The upper limit of the *FF* values (i.e., the yellow curve) can be explained by the interplay between charge recombination and extraction processes[27,28]. From the perspective of the active layer materials and their morphology, the carrier lifetime and mobility are the most important characteristics, where one of the effective ways to analytically describe the characteristics is using the electronic quality $Q$ ($Q = \frac{\mu_e \mu_h}{k_2}$, where $\mu_e$ is the electron mobility, $\mu_h$ the hole mobility and $k_2$ the bimolecular recombination coefficient)[27]. We calculated the $Q$ values with variable $FF$ and voltage loss by fixing the bandgap at 1.4 eV, the external quantum efficiency at 80%, and the thickness at 100 nm. For $Q$ values, we extract the relation between $FF$ and voltage loss, and show the results in **Supplementary Figure 2**. When the $Q$ value is large (e.g., 1000 cm$V^{-2}s^{-1}$), where the transport and recombination losses are minimized, the curve shows that a small voltage loss corresponds to a large $FF$. This conforms to the upper limit of the $FF$ values, because the limit converges to the ideal solar cells described in **Figure 1a**. That is, the decrease in $FF$ is due to the decrease in $V_{OC}$ and the geometric change in the exponential curves. When the $Q$ value decreases, the $FF$ curve shifts downwards because transport loss or recombination begins to dominate the $FF$. This shows that $Q$ plays a vital role in explaining the scattering of the $FF$ values from the perspective of the fundamental properties of semiconductors.

In addition to $Q$ (material properties), device fabrication and operation can also influence charge extraction and charge recombination. For example, the geometry of a device and its layers, the contacts between layers, and the illumination intensity influence charge extraction and recombination. Hence, a dimensionless figure of merit (FoM) $\alpha$ (refer to **Supplementary Figure 3** and related notes for details) can be used to account for both materials- and device-related effects.[29] A device with a small $\alpha$ (typically <1) means that the device has an $n_{id}$ close to 1, while a large $\alpha$ means that it is transport-limited. By fixing the optical gap, we can calculate the relationship between $FF$ and $V_{OC}$ loss at a set of $\alpha$ values. While more parameters were considered by introducing $\alpha$ values, the $FF$-voltage loss curves share a similar trend as the cases where only $Q$ values are considered in the earlier paragraph. This trend shows that under the condition of similar charge extraction and recombination, the $FF$ values increase with decreasing voltage losses, which is consistent with the $FF$ limit in **Figure 1b**. However, the reason for the left border of the data (as shown by the green curve), which shows a rapid decrease in $FF$ with decreasing voltage losses, is unclear.

To understand the competition between $FF$ and $V_{OC}$ that results in the left border of the scattered data in **Figure 1b**, we selected four donor-acceptor combinations at the left border of the data points. They are PM6:Y11, PM6:Y1, PM6:IEICO-4F and PTO2:Y1, with $V_{OC}$ losses of 0.53 V to 0.49 V and $FF$ 0.65 to 0.27 (device characteristics shown in **Supplementary Figure 4**). We first analyse the mobility of the blends by fabricating single-carrier devices (**Supplementary Figure 5**). We find that all of them have electron and hole mobilities of approximately $10^{-4}$ cm$^2$v$^{-1}$s$^{-1}$, consistent with previous reports, where these polymer donors and NFAs have been demonstrated to possess good transport properties and can achieve a PCE above 10% with an *FF* of approximately 70%[5,30,31]. The low-voltage-loss OSC blends in this work do not have particularly low mobilities. We then obtain the recombination coefficients of the blends using the time-delayed collection field (TDCF) method and show that the bimolecular recombination coefficient is high in the low-voltage-loss cases (see $k_2$ in **Supplementary Table 2**), agreeing with previous reports[32]. The reason for the high $V_{OC}$ in these cases can result from a relatively high emission quantum yield of the bimolecular recombination[3,4]. Using all these parameters (**Supplementary Table 2**) derived from the measurements, we can calculate the theoretical *FF* values using Supplementary Equation (5)-(7). Interestingly, as shown in **Table 1**, the PTO2:Y1 case shows a reconstructed *FF* significantly higher than the experimental value, indicating that increased bimolecular recombination alone cannot rationalise the decreased *FF*. In other words, other factors which are not included in $Q$ and $\alpha$ affect the *FF*.

## Fill factor and free charge generation

As shown in **Figure 1c**, OSCs are based on intrinsic excitonic materials. The generation and recombination of free charges (FC) in OSCs relies on intermediate states. Compared with inorganic semiconductors, the decay of excitons (Ex) or donor-acceptor-charge-transfer (DA-CT, hereafter CT) states to the ground state and the back-transfer from CT states to Ex can lead to geminate recombination. Thus, we hypothesise that geminate recombination processes, which are intrinsic to OSCs, can additionally limit the FC generation and affect the *FF* in OSCs.

To investigate the relationship between geminate recombination and *FF* in the selected systems, we measured FC generation via TDCF. By adjusting the delay time and applying a large reverse bias, we can measure the amount of FC carriers generated at a set of pre-biases. In contrast to the photocurrent, the FC generation data from TDCF are not influenced by the non-geminate loss because all the carriers are extracted in this case. As illustrated schematically in **Supplementary Figure 6**, the gap between the photocurrent and FC generation originates from the free carriers that recombine before extraction. This loss exists in all kinds of solar cells and is due to transport and recombination, which can be quantified by the FoM $\alpha$. The gap between the generated FC and absorbed photons is lost due to geminate recombination. To obtain the flux of the absorbed photons and display geminate recombination in the active layers, we performed ellipsometry measurements and calculated the light distribution of the different layers in the device (**Supplementary Figure 7**). Combining the photocurrent and FC generation data, we show the TDCF experimental results in **Figure 2a-d**, where the pre-bias is swept from 0 to near $V_{OC}$. The dashed blue lines stand for the number of extracted charges, i.e., the current density corresponding to FC generation ($J_G$). The reddish area below the dashed blue line stands for loss due to geminate recombination. In all four material examples, this part is dependent on the voltage biases, showing that geminate recombination affects the *FF* in our OSCs.

To analytically describe the influence of FC generation on the *FF*, we use the relationship between current density and voltage in OSCs[29],

$$J(V) = J_G(V)\left\{\exp\left[\frac{q}{k_B T(1+\alpha)}(V - V_{OC})\right] - 1\right\} \quad (3)$$

where $J(V)$ is the current density at voltage $V$, $J_G$ is the current density corresponding to FC generation, $J_0$ is the reverse saturation current, $q$ is the elementary charge, $k_B$ is the Boltzmann constant, $T$ is the temperature of the device, and $\alpha$ is the FoM. To further parameterise the field-dependent FC generation, we apply an analytical analysis with detailed derivation in **Supplementary Note 1**, which will be turned into a more advanced model later. The FC generation derived is

$$J_G(V) = J_{Abs}\eta_{int}\left[1 + \left(\frac{1}{\eta_{int}} - 1\right)\beta|V - V_{OC}|\right] \quad (4)$$

where $J_{Abs}$ is the generation current density due to optical absorption, $\eta_{int}$ is the internal generation efficiency on the open-circuit (OC) condition, $\beta$ is the field-dependence coefficient with unit V$^{-1}$ which physically represents the influence of an electric field on the geminate recombination term $\left(\frac{1}{\eta_{int}} - 1\right)$. $\beta$ can be deduced together with $\eta_{int}$ from the FC generation experiments (e.g., TDCF). We simulate the JV curves of PM6:Y11, PM6:Y1, PM6:IEICO-4F, and PTO2:Y1 as shown in **Figure 2e**. The parameters for the simulation are listed in **Supplementary Table 3**, and were derived from the TDCF and ellipsometry measurements. The simulation curves reproduce the JV curves in **Figure 2a-d**, indicating that Equation (4) captures the key physical parameters that determine the *FF*.

We systematically vary the parameters, as shown by the JV curves in **Supplementary Figure 8**, to quantify the effects of different parameters (that is, FC generation efficiency, charge transport, and field dependence of FC generation) on *FF* in **Figure 2f.** In the case where FC generation is dependent on the field ($\beta > 0$), the *FF* decreases with decreasing $\eta_{int}$, with *FF* significantly affected by small $\eta_{int}$ and/or large $\beta$ values. We have also included the four NFA examples as dots in **Figure 2f**. PM6:Y11 and PM6:Y1 have efficient FC generation (large $\eta_{int}$), in the region where the field-dependent FC generation has a small effect on the *FF*. While the FC generation in PM6:IEICO-4F is not as efficient as PM6:Y11 or PM6:Y1, the FC generation is least dependent on the field among all the samples (with a small $\beta$ value of 0.05 V$^{-1}$), also resulting in a small effect of the FC generation on the *FF*. Consequently, the *FF* values of PM6:Y11, PM6:Y1, and PM6:IEICO-4F are mainly regulated by their transport properties. In contrast, PTO2:Y1 has a small FC generation efficiency ($\eta_{int} = 0.06$) and a moderate field-dependence $\beta$ of 0.09 V$^{-1}$, resulting in a small *FF* value dominated by the field-dependent FC generation.

The significant drop in *FF* in PTO2:Y1, which is limited by field-dependent FC generation, presents an *FF* limit to the left part of **Figure 1b**. To shift this boundary further towards the left (*i.e.,* achieving a high *FF* at low voltage losses for enhanced power conversion efficiencies), we need to understand the physical nature of field-dependent FC generation, which we aim to elaborate on in the next section.

## Fill-factor limit and geminate pairs

To figure out the nature of the primary bound state limiting the *FF*—whether it is the transition from the Ex to CT states or from CT states to FC states—we applied the pump-push photocurrent (PPPC) measurements (**Supplementary Figure 9**). This technique utilises an IR 'push' pulse to selectively probe the dynamics of bound states which do not naturally contribute to photocurrent, enabling us to identify the dynamics of geminate losses[33,34] and the binding energies of the key states involved. The geminate pairs in our blends were found to have similar decays as the species in the neat-acceptor films, indicating that the geminate loss here was due to the Ex of acceptors. The extracted binding energies for the key bound states in blends in pure acceptor films were also similar. This indicates that the dissociation of acceptor excitons in the blends is probably the key step assisted by the built-in voltage and affecting the *FF* of the devices.

To monitor the Ex population under an applied voltage, we utilized bias-dependent PL, a productive approach for monitoring electronic states in devices[35–38]. In inorganic solar cells, PL typically arises from band-to-band free carrier recombination and is easily quenched by charge extraction, as shown in **Supplementary Figure 10** for silicon solar cells. In OSCs, The PL intensity remains strong even under SC or negative-bias conditions because the PL primarily originates from geminate recombination rather than non-geminate processes, consistent with previous reports[37,39]. Our PPPC results also suggest that the PL in the studied systems is mainly due to Ex recombination rather than CT recombination. Therefore, PL intensity serves as a reliable indicator of Ex population and decay dynamics in these OSCs. Here, the Ex may have a complex density including local excitons as well as more delocalized intermolecular states observed in some crystalline molecular acceptors including Y6[40,41].

To reveal the factors that influence geminate recombination, we systematically selected several D:A combinations and measured their bias-dependent PL. Here, we studied two representative polymer donors (PM6 and PBDB-T) and three NFAs (IT-4F, Y11, and Y1) to obtain blends with different energetic offsets. Here, energetic offsets PM6, also known as PBDB-T-2F or PBDB-TF, contains two fluorine atoms substituted on the conjugated side groups of the BDT units. Fluorination leads to a higher ionisation energy (IE) for PM6 than for PBDB-T[42,43]. Therefore, blends with PM6 have smaller IE offsets when paired with the same acceptor. Compared to IT-4F, benzotriazole was introduced into the core, and cyclopentadienyls were substituted with pyrrole rings for Y1 and Y11. For the end groups, Y11 and IT-4F were fluorinated, whereas Y1 was not. The IE decreases from Y1 to Y11, and then to IT-4F[5,31,44]. We integrated the PL emission intensity and plotted it as a function of the electric field (see **Supplementary Figure 11** for the PL spectra) (**Figure 3a-g**), where we also show the current densities. The electric field ($F$) is translated from the applied voltage bias ($V$) as $F = \frac{V - V_{OC}}{d}$, where $d$ is the film thickness of the active layer. For comparison, the figures in a row have the same donor and the figures in a column have the same acceptor.

We compared the effects of the electric field on the PL intensities of these blends with different energetic offsets. In **Figure 3a** and **b**, the PL intensities of the PM6 blends with Y1 and Y11 show obvious drops with increasing electrical fields. This is consistent with the field-dependent FC generation results presented in our previous discussions of these two blends, showing that the dissociation of unseparated charges can be facilitated by an external voltage bias. In contrast, the bias dependence of the PL of the PBDB-T-based blends (**Figure 3d and e**) is not as significant as that of the PM6 based blends, suggesting that the high-offset system exhibits low field-dependent Ex dissociation. This is also demonstrated in PM6:IT-4F and PBDB-T:IT-4F (**Figure 3c and f**), which show the smallest field-dependent PL and the largest IE energetic offsets (IT-4F has the lowest IE compared with the others). In conjunction with the photocurrents, the field-dependent PL of these blends show a negative correlation between the energetic offset and field-dependent Ex dissociation process. Here, Ex dissociation refers to the full process from the Ex to FC states, which is typically linked by the CT states. Based on PPPC and field-dependent PL spectroscopy, we believe that the electric field mainly influences the transition between the Ex and CT states, which is consistent with the findings in a recent report[37]. Despite many discussions on the CT dissociation (Process 5, **Figure 1c**) with electric fields[45–49], the field dependence of CT dissociation is likely to be a minor factor in current NFA systems. In transient absorption spectroscopy, the signal of FC appears at the same time as the charge transfer between a donor and an acceptor occurs[40,50]. This indicates an efficient transition process from the CT states to the FC states if the CT state is an intermediary between the Ex and FC states, which reduces the possibility that field-dependent Ex dissociation is via an equilibrium shift due to the CT dissociation. Another possible approach involves the field-assisted direct separation of the Ex into the FC. However, the PL intensity of the pristine material films is field-independent within our investigation range, as shown in **Supplementary Figure 11g and j**, indicating that direct dissociation is unlikely. A recent report showed similar conclusions by blending the NFA Y5 with inert polystyrene[36]. Therefore, since we have excluded other possibilities, the transition between the Ex and CT states is the most probable candidate for being affected by the electric field. Back to the field-dependent PL of blends, the electric field acting on the Ex-CT transition explain the negative correlation between field-dependent Ex dissociation process and the energetic offsets, because the energetic offsets are directly linked to the Ex-CT transition[11,37]. We do not exclude the possibility that the FC can be formed via a long-range CT[51]. Hence, we can conclude that the electric-field-dependent transition between the Ex and CT states is the primary mechanism of the field-dependent FC generation, which limits the *FF* in OSCs.

## Fill-factor limit and donor-acceptor charge transfer

While the field-assisted dissociation of electron-hole pairs[52–55] and CT states in OSCs[48,56–59] have been studied, a unified description of how the energetic offset and applied voltage influence the *FF* is still lacking.

To quantitatively elucidate the effect of the field-dependent Ex dissociation process and its relationship with the *FF* of OSCs, we develop an analytical model. The transition rates are achieved with the Marcus theory and the electrostatic potential is incorporated using perturbation theory to account for the first- and second-order Stark effect (see **Supplementary Note 2** for detailed description). In short, the energy of electronic states is shifted by the external electric field due to their electric dipole and polarizability, leading to the electric-field-dependent rate coefficient $k_{Ex-CT}$ and $k_{CT-Ex}$, as shown in **Figure 4a**. As the electric field increases, $k_{Ex-CT}$ increases and $k_{CT-Ex}$ decreases. Thus, the field-dependence coefficient $\beta$ in equation (4) can be attributed to the electric dipole and polarizability of CT states, while the polarizability is more important in a uniform bulk heterojunction. Using these rate coefficients, the complete model (**Figure 1c**) is integrated into a device-level drift-diffusion simulation, as described in **Supplementary Note 3**. By decreasing the energy offset, we reproduce the JV curves with a reduced *FF* (**Supplementary Figure 12**). As exciton dissociation also competes with exciton decay, a slow exciton decay rate can significantly mitigate this field dependence. As shown in **Figure 4b**, although the *FF* inevitably decreases at lower voltage losses, a slower exciton decay rate preserves a high *FF* down to a smaller voltage loss. It can also be understood from the perspective of detailed balance, that the Ex dissociation does play an important role in the equivalent recombination and generation of FC[59]. This finding is significant as it reveals a pathway to simultaneously improve both the *FF* and $V_{OC}$ by increasing the Ex lifetime, and the conclusions can further be extended to other factors that limit Ex splitting, e.g., reorganization energy and exciton diffusion.

The conclusion that suppressing exciton recombination is crucial for a high *FF* aligns well with the developmental trajectory of OSCs over the past decade. To illustrate this trend, we grouped the data points from **Figure 1b** by their publication year and replotted them **Figure 4c**. The groups represent distinct material eras: reports before 2015 (orange dots), reports between 2015 and 2019 dominated by ITIC and its derivatives (blue dots), and reports after 2019 following the development of Y6 (green dots). We clearly see a higher *FF* and lower voltage losses when moving from ITIC-based OSCs to Y6-based ones. One of the key features of Y6-based NFAs is their slower Ex decay rate compared with that of ITIC-based NFAs (**Supplementary Figure 13**, refer to **Supplementary Table 4** for details[11,60,69–71,61–68]). Despite the possibility of delocalized intermolecular states in neat acceptors, they would be coupled to the emissive states at room temperature[72]. In our time-correlated single-photon counting (TCSPC) measurements performed on Y6, L8-BO, ITIC and IT-4F samples (**Supplementary Figure 14**), Y6 and L8-BO show emissive-state lifetime of approximately three times longer than that of the ITIC family. However, Y11 and Y1 show relatively short lifetimes compared to Y6 and L8-BO, a finding consistent with their lower *FF* and field-dependent FC generation as shown in **Figure 2** and **3**.

Our results have important implications for the rational material design of OSCs for the next breakthroughs in efficiency. In addition to the established understanding of charge transport, which is decisive for the *FF* of all solar cells, the effect of field-dependent FC generation on *FF* has been elaborated in this work. In terms of utilizing or suppressing the field-dependent FC generation, there is room for further exploration and optimization in the electric dipole orientation and the polarizability according to the first- and second-order Stark effect by manipulating donor-acceptor configurations. Compared with previous works on FC generation efficiency[11,37], we further highlight that long Ex lifetime can also help us to better design materials for high $V_{OC}$ and high *FF*. This insight resolves a critical design question for minimizing voltage loss: pursuing a high radiative recombination rate to achieve a high photoluminescence quantum yield is ultimately counterproductive, as the resulting short exciton lifetime is detrimental to the *FF*. Therefore, future material design should prioritize the suppression of non-radiative recombination or find a way to utilize field-dependence to simultaneously enhance $V_{OC}$ and *FF*.

## Conclusion

In summary, this work provides a comprehensive framework for understanding the relationship between the *FF* and voltage loss in OSCs. Although we acknowledge that the transport limit remains an important factor in OSCs, we demonstrate that field-dependent FC generation also limits the *FF*, a mechanism of particular significance in OSCs. We have shown both experimentally and theoretically that this field dependence originates from the transition between Ex and CT states. Our experimental results reveal its connection to the energetic offset and voltage bias, while our theoretical model confirms this by incorporating the Stark effect in donor-acceptor blends. Consequently, we conclude that limitations to the *FF* in OSCs stem not only from non-geminate recombination but also, critically, from geminate recombination. Our work reveals that enhancing exciton dissociation is a key strategy for further increasing the *FF* and the efficiency of OSCs.

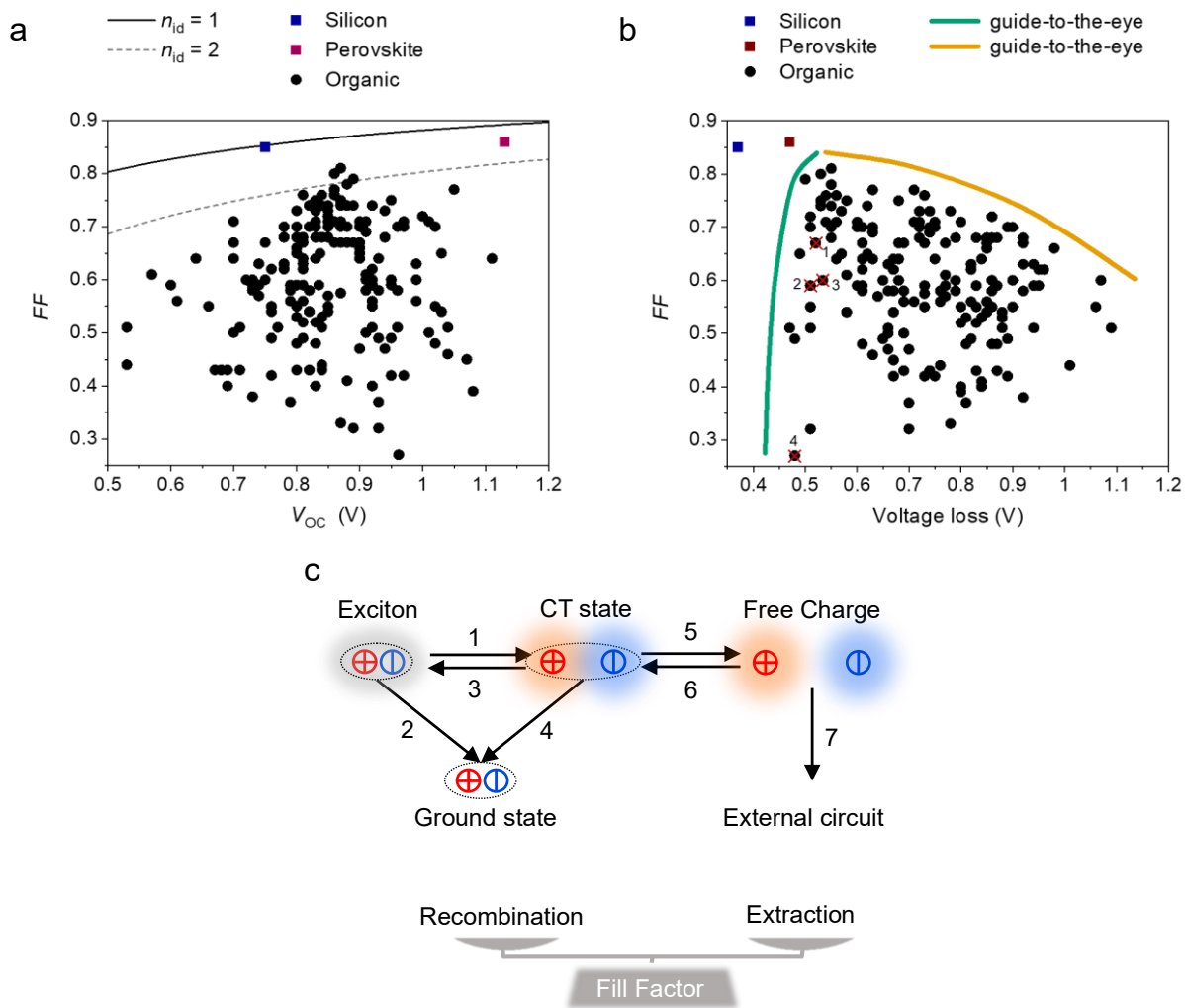

**Figure 1 | Fill factor (*FF*) of organic solar cells (OSCs). a** The *FF* limit as a function of $V_{OC}$ for single-junction solar cells is illustrated by the solid curve ($n_{id}$ = 1) and the dashed curve ($n_{id}$ = 2). The blue square represents a silicon solar cell. The red square represents a perovskite solar cell. The black dots represent OSCs. **b** The $V_{OC}$ data is converted to voltage losses for allowing a band-gap independent comparison. Points 1 to 4, marked with a red cross, are PM6:Y11, PM6:Y1, PM6:IEICO-4F and PTO2:Y1, respectively. **c** An illustration of the charge dynamics in OSCs. Process 1-6 form effective charge recombination which competes with charge extraction which is process 7.

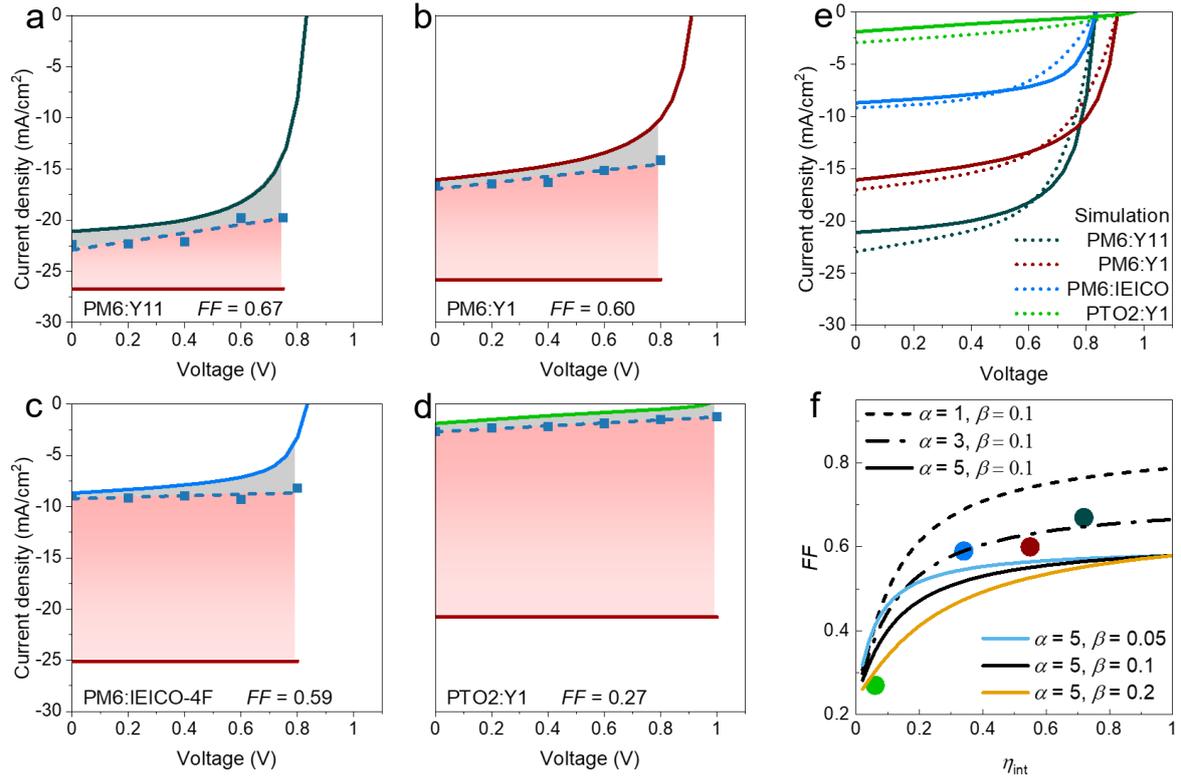

**Figure 2 | Field-dependent charge generation and influence on *FF*. a-d** The charge generation of PM6:Y11, PM6:Y1, PM6:IEICO-4F and PTO2:Y1. The upper solid curves are the current-voltage density (JV) measured under one-sun conditions. The light blue squares are free carriers measured via TDCF. The dashed light blue lines represent the charge generation current densities ($J_G$) obtained by linearly fitting the dissociated charges. The red lines represent photons absorbed by the device ($J_{Abs}$). The grey areas represent the non-geminate recombination losses. The red areas represent geminate recombination losses. **e** Solid curves are measured JV curves corresponding to the four examples. The dashed curves are simulated JV curves which show the influence of the field dependence on *FF*. **f** *FF* limit as a function of internal FC generation efficiency on the open-circuit condition is simulated with different transport limit ($\alpha$) and different field dependence ($\beta$). The solid-coloured circles are PTO2:Y1, PM6:IEICO-4F, PM6:Y1, and PM6:Y11 from left to right.

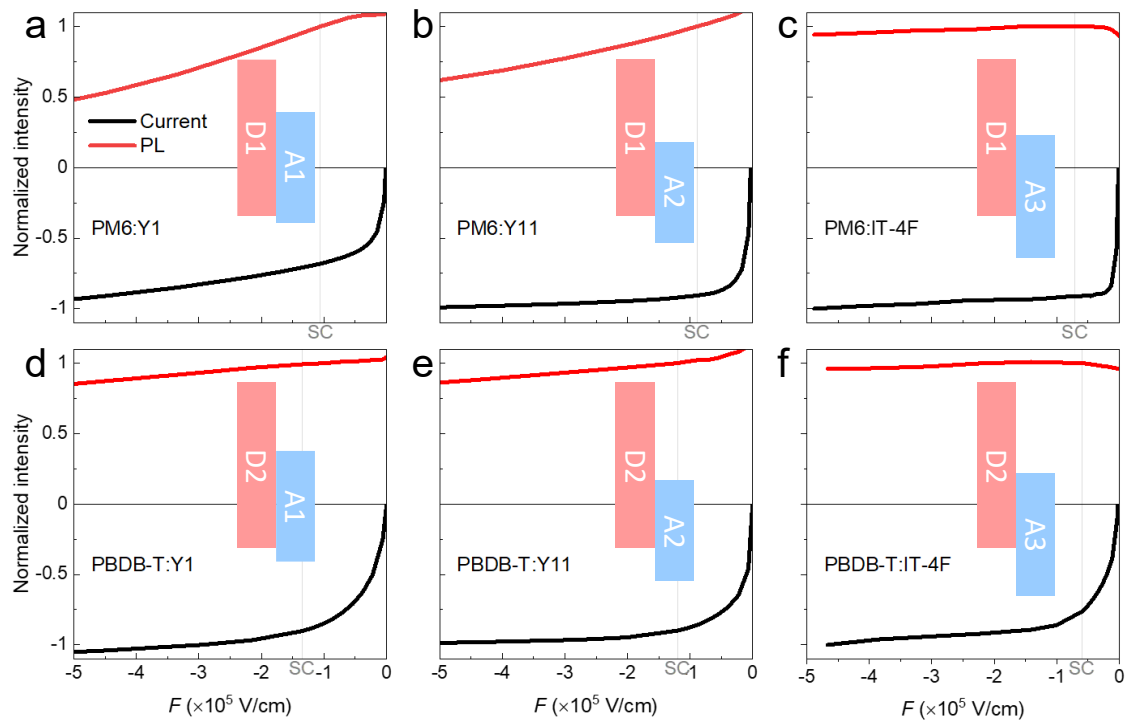

**Figure 3 | Bias-dependent photoluminescence.** Figures in the same row show data from the same donor and figures in the same column show data from the same acceptor. The black curves are the PL intensities obtained by integrating the counts of the acceptor emission peaks and are normalised to the values at SC. The red curves are normalized JVs which are measured simultaneously with the PL. The pink and blue rectangles are schematic representations of the relative energy levels between the donor (D) and the acceptor (A).

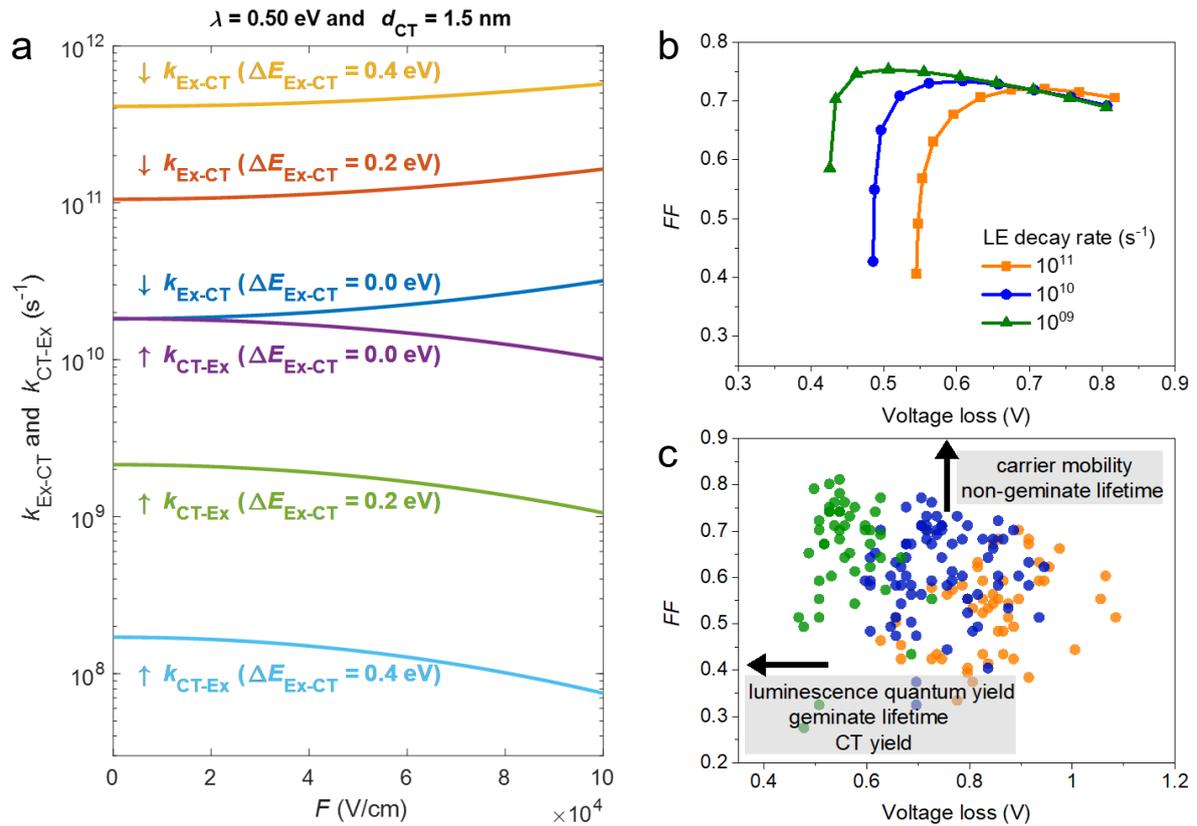

**Figure 4 | Potential for further improvements in *FF* and OSC performance. a** Simulated transition rate between the Ex and CT states dependent on the internal electric field due to the Stark effect. **b** Simulated *FF*-voltage loss limit from the Ex to CT states at different Ex decay rate. Dots with the same colour have energy offsets varying from 0.05 eV to 0.45 eV. **c** Green dots are based on NFA reports after 2019, when Y6 was developed. Blue dots are NFAs reported between 2019 and 2015, when ITIC and their derivatives dominated the field. Orange dots are NFA reports before 2015. The two shaded arrows represent two routes to further improve the performance of OSCs by increasing *FF* and $V_{OC}$ simultaneously.

|  | **PM6:Y11** | **PM6:Y1** | **PM6:IEICO-4F** | **PTO2:Y1** |
| --- | --- | --- | --- | --- |
| **Reconstructed *FF*** | 0.66 | 0.61 | 0.56 | 0.58 |
| **Measured *FF*** | 0.67 | 0.60 | 0.59 | 0.27 |

**Table 1 | Comparison of *FF*.** Reconstructed *FF* is based on measured transport and recombination parameters with Supplementary Equation (5)-(7). The device based on PTO2:Y1, with low FC generation, shows a lower reconstructed *FF* compared with the measured *FF*.

# Experimental methods

**Material preparation**: The electron transport material (ZnO-N10, nanoparticle solution) was purchased from Avantama AG and used without additional treatment. The hole transport material (molybdenum oxide, powder) was purchased from Sigma-Aldrich and used without additional treatment. Among the active layer materials, PM6, PTO2, and IEICO-4F were purchased from Solar Materials, Inc. (Beijing). Y1 and Y11 were synthesised at the Central South University.

**Fabrication of photovoltaic devices:** Pre-patterned indium tin oxide substrates were cleaned with detergent followed by two 20-min ultrasonic steps in acetone and isopropanol. Subsequently, a 10-min ultraviolet–ozone treatment was applied. A layer of zinc oxide (N-10, Avantama) of approximately 30 nm thickness was spin-coated in air at 3600 rpm and annealed at 120 °C for 5 min, after which the samples were moved into a glove box. The active layer was spin-coated from the solution, and the rotation speed was adjusted to yield an active layer thickness of around 100 nm. Chloroform was used as the solvent for PM6:Y11, PM6:Y1 and PTO2:Y1. Chlorobenzene was used for PM6:IEICO-4F. The ratio and total concentration were 1:1.2 and 18 mg ml$^{-1}$ for PM6:Y11, 1:1 and 16 mg ml$^{-1}$ for PM6:Y1 and PTO2:Y1, and 1:1 and 20 mg ml$^{-1}$ for PM6:IEICO-4F. The solution was kept on a hot plate at 60 °C for 12 h before spin coating and was kept on a plate during spin coating. Immediately after spin-coating, an annealing step at 100 °C for 10 min was applied. The substrate was then placed on a mask and transferred to a vacuum chamber. Molybdenum oxide (12 nm) and Ag (200 nm) were thermally evaporated in a vacuum of approximately 10$^{-6}$ mbar.

**JV characterisation:** Devices were encapsulated in a glovebox and measured in air. The active area of the tested solar cell was 4 mm$^2$. The J-V curves (measured in the forward direction, that is, from negative to positive bias, with a scan step of 0.04 V) were collected by using a Keithley 2400 Source Meter under AM1.5 illumination provided by a solar simulator (LSH-7320 ABA LED solar simulator) with an intensity of 1000 W m$^{-2}$ after spectral mismatch correction. The light intensity for the J-V measurements was calibrated using a reference Si cell (VLSI standards SN 10510-0524 certified by the National Renewable Energy Laboratory).

**EQE$_{PV}$ measurements**: EQE measurements were conducted using an integrated system, QE-R3011). The system was calibrated by using a silicon reference detector.

**PL and field-dependent PL measurements:** Spectrally resolved PL measurements were performed on an Andor Shamrock-303i spectrograph equipped with an Andor Newton EM-CCD (DU970N-UVB). During the measurement, the CCD detector was cooled to -45 °C. The wavelength of the system was calibrated by using a mercury lamp. The intensity of the spectra was calibrated using a standard halogen lamp (AvaLight-HAL-S-Mini, Avantes). The PL was excited by a Thorlabs collimated laser-diode-pumped DPSS laser module (CPS532), and a 550 nm long-pass filter was used to protect the detector. The PL excitation and detection were performed using an objective, a 45 °-placed 552 nm long-pass dichroism glass, and optical fibres. The samples were positioned where the light spot area was larger than the device pixel size. The laser intensity was controlled by a neutral-density filter wheel to ensure that the photocurrent was equal to that of one sun. A source meter Keithley 2400 was connected to the photovoltaic devices to apply a voltage bias and record the current response.

**EQE$_{EL}$ measurement**: The EQE$_{EL}$ was recorded using a home-built system with a Hamamatsu silicon photodiode 1010 B. A Keithley 2400 was used for supplying bias voltages and recording injected current, and a Keithley 485 was used for collecting the photo-current generated from the emitted photons of the samples.

**TDCF measurement**: TDCF investigations were performed with an Agilent Technologies DS05054A oscilloscope using 50 Ω input resistor and a Tektronix AFG 3101 function generator. The samples were excited by radiation from the optical parametric amplifier Topas-C (Light Conversion Ltd.) pumped by a femtosecond Ti:sapphire laser Integra-C (Quantronix Inc. generating 130 fs duration pulses at a 430 Hz repetition rate. linear optical parametric amplifier TOPAS-C was used to generate an excitation pulse emitting at 515 nm.

**Ellipsometry measurement and analysis**: Ellipsometry was performed with a Mueller Matrix Ellipsometer (RC2, J.A. Woollam Co., Inc). CompleteEASE was used to globally fit the Mueller matrix data with the B-spline and General Oscillator models for the optical properties of the samples. The refractive index and distinction coefficient were used for transfer matrix optical modelling (https://github.com/erichoke/Stanford/tree/master).

**TCSPC measurement**: TCSPC was operated using a system from Edinburgh Instruments with a microchannel plate photomultiplier tube (Hamamatsu). An excitation pulse laser at 405 nm was generated using a pulsed picosecond diode (Hamamatsu).

**PPPC measurements**: fs pulses (800 nm, 35 fs) were generated using a 4 kHz Ti:sapphire regenerative amplifier (Astrella, Coherent). These pulses were routed onto two optical parametric amplifiers (TOPAS Prime, Coherent). The 1400 nm output from one TOPAS passed through a frequency-doubling barium borate crystal to generate the pump at 700 nm. The pump pulse was then directed onto a mechanical delay stage to vary the time delay between the pump and the push beams. The 2000 nm output from the other TOPAS serves as the push. The push was mechanically modulated at 1.1 kHz. Both pump and push pulses were aligned to a single spot on the device pixel. During the measurements, the devices were connected to a lock-in amplifier (MFLI, Zurich Instruments) and measured under short-circuit conditions. The reference current, $J$, was measured at a pump frequency of 4 kHz, and the push-induced current, $\Delta J$, was measured at a push frequency of 1.1 kHz. The devices were placed in a cryostat (HFS600E-PB4, Linkam) with a liquid-nitrogen cooling module (LNP96, Linkam) to control the temperature.

# Acknowledgement

The authors thank O. Inganäs, J. Hou, J. Durrant, N. Jain, R. Zhang, X. Zhou, D. Qian and Y. Wang for insightful discussions. The authors thank M. Azzouzi for helpful discussions and for making the DriftFusionOPV code publicly available. The drift-diffusion simulations presented in this work were performed using this code, developed in the J. Nelson Group. H.Z. acknowledge the financial support from the China Scholarship Council (grant no. 201706100186). The research in Linköping was supported by the Swedish Strategic Research Foundation through a Future Research Leader program to F.G. (FFL 18-0322) and the Swedish Government Strategic Research Area in Materials Science on Functional Materials at Linköping University (faculty grant no. SFO-Mat-LiU #2009-00971). F.G. is a Wallenberg Scholar. D.N. acknowledges funding by the Deutsche Forschungsgemeinschaft (DFG, German Research Foundation) through the project Extraordinaire (Project Number 460766640).

## Author contributions

H.Z. and F.G. conceived the ideas. H.Z. designed the project, prepared devices and samples, performed JV characterizations, $EQE_{PV}$ measurements, PL and field-dependent PL measurements, $EQE_{EL}$ measurements, ellipsometry measurements and analysis. J.Y. synthesized the Y1 and Y11 acceptor materials under the supervision of Y.Z.. T.W. and H.Z. performed the PPPC measurements under the supervision of A.A.B.. R.J. performed the TDCF measurements under the supervision of V.G.. Y.Liu performed the TCSPC measurements under the supervision of X.Z. Y.Li prepared the samples for PPPC measurement. N.T. helped with the initial calculation of charge transfer. H.Z. developed the theoretical model with the help of T.K., V.C. and D.N.. D.N. supervised the simulation. F.G. supervised the project. H.Z. and F.G. wrote the manuscript. All authors discussed the results and commented on the final manuscript.

## Competing interests

The authors declare that they have no competing interests.

Supplementary Information

# Understanding the fill-factor limit of organic solar cells

## Contents

| Donor | Acceptor | $E_{opt}$ (eV) | $V_{oc}$ (V) | $V_{oc}$ loss (V) | FF (%) | $J_{sc}$ (mA/cm²) | PCE (%) | FFn | Thickness (cm) | Q | Ref |
|---|---|---|---|---|---|---|---|---|---|---|---|
| PBTI3T | Hexyl-PDI | 1.88 | 1.08 | 0.8 | 0.39 | 1.51 | 0.65 | -0.53 | 6.00E-06 | 1.44E-02 | 73 |
| | Phenethyl-PDI | 1.88 | 1.02 | 0.86 | 0.48 | 2.44 | 1.2 | -0.23 | 6.00E-06 | 7.15E-02 | |
| | Phenyl-PDI | 1.88 | 1.02 | 0.86 | 0.55 | 6.56 | 3.67 | 0.01 | 6.00E-06 | 2.68E-01 | |
| PBDB-TS1 | PPDI | 1.61 | 0.8 | 0.81 | 0.53 | 12.85 | 5.45 | -0.03 | 8.80E-06 | 2.05E+00 | 74 |
| PBDTTT-C | PPDI | 1.65 | 0.76 | 0.89 | 0.49 | 10.46 | 3.78 | -0.16 | 8.80E-06 | 1.29E+00 | |
| PBDT-TS1 | H-tri-PDI | 1.63 | 0.73 | 0.9 | 0.6 | 16.52 | 7.25 | 0.23 | 8.00E-06 | 4.96E+00 | 75 |
| PBDTT-F-TT | di-PBI | 1.67 | 0.8 | 0.87 | 0.59 | 11.98 | 5.9 | 0.18 | 1.05E-05 | 7.56E+00 | 76 |
| PDBT-T1 | SdiPBI-S | 1.88 | 0.9 | 0.98 | 0.66 | 11.98 | 7.16 | 0.40 | 1.10E-05 | 1.62E+01 | 77 |
| PDBT-T1 | SdiPBI-Se | 1.86 | 0.96 | 0.9 | 0.7 | 12.49 | 8.42 | 0.52 | 1.10E-05 | 2.50E+01 | 78 |
| PBDTTT-C-T | PDI1 | 1.7 | 0.85 | 0.85 | 0.54 | 8.86 | 3.91 | 0.00 | 1.00E-05 | 2.58E+00 | 79 |
| PBDTTT-C-T | Bis-PDI-T-EG | 1.7 | 0.97 | 0.73 | 0.42 | 0.33 | 0.1 | -0.43 | 1.00E-05 | 8.06E-02 | |
| P3TEA | SF-PDI2 | 1.72 | 1.11 | 0.61 | 0.64 | 13.27 | 9.5 | 0.30 | 1.20E-05 | 1.02E+01 | 8 |
| PffBT4T-2DT | SF-PDI2 | 1.72 | 0.96 | 0.76 | 0.58 | 11.04 | 6.1 | 0.12 | 1.20E-05 | 6.46E+00 | |
| P3TEA | diPDI | 1.72 | 0.95 | 0.77 | 0.59 | 12.46 | 7 | 0.15 | 1.20E-05 | 7.94E+00 | |
| PffBT4T-2DT | diPDI | 1.72 | 0.84 | 0.88 | 0.53 | 11.19 | 4.9 | -0.04 | 1.20E-05 | 5.01E+00 | |
| PBDTTT-C-T | S(TPA-PDI) | 1.65 | 0.87 | 0.78 | 0.33 | 11.27 | 3.22 | -0.73 | 1.00E-05 | 1.18E-01 | 80 |
| PTB7-Th | Ph-PDI | 1.65 | 0.85 | 0.8 | 0.55 | 11.91 | 5.15 | 0.03 | | | 81 |
| PTB7-Th | Ta-PDI | 1.65 | 0.78 | 0.87 | 0.68 | 17.1 | 8.91 | 0.50 | | | |
| PffBT-T3(1,2)-2 | TPPz-PDI 4 | 1.7 | 0.99 | 0.71 | 0.56 | 12.5 | 6.9 | 0.04 | 1.00E-05 | 2.66E+00 | 82 |
| PffBT-T3(1,2)-2 | TPE-PDI 4 | 1.7 | 1.03 | 0.67 | 0.54 | 10.6 | 5.9 | -0.03 | 1.00E-05 | 1.77E+00 | |
| PffBT-T3(1,2)-2 | TPC-PDI 4 | 1.7 | 1.04 | 0.66 | 0.51 | 8.7 | 4.6 | -0.13 | 1.00E-05 | 1.10E+00 | |
| PTB7-Th | TPB | 1.65 | 0.79 | 0.86 | 0.58 | 18.4 | 8.47 | 0.15 | 8.00E-06 | 3.31E+00 | 83 |
| PTB7-Th | Helical PDI 1 | 1.73 | 0.79 | 0.94 | 0.59 | 11 | 5.14 | 0.18 | 9.00E-06 | 4.37E+00 | 84 |
| PBDTT-TT | Helical PDI 1 | 1.7 | 0.8 | 0.9 | 0.55 | 13.5 | 5.94 | 0.04 | 1.30E-05 | 1.04E+01 | |
| PTB7 | hPDI3 | 1.72 | 0.76 | 0.96 | 0.62 | 13 | 6.3 | 0.29 | 1.00E-05 | 1.12E+01 | 85 |
| PTB7-Th | hPDI3 | 1.66 | 0.81 | 0.85 | 0.67 | 14.3 | 7.7 | 0.45 | 1.00E-05 | 1.97E+01 | |
| PTB7 | hPDI4 | 1.72 | 0.78 | 0.94 | 0.63 | 12.9 | 6.4 | 0.32 | 1.00E-05 | 1.18E+01 | |
| PTB7-Th | hPDI4 | 1.66 | 0.8 | 0.86 | 0.68 | 15 | 8.1 | 0.49 | 1.00E-05 | 2.44E+01 | |
| PTB7-Th | PDI-T | 1.72 | 0.88 | 0.84 | 0.41 | 9.2 | 3.52 | -0.45 | 9.00E-06 | 3.18E-01 | 86 |
| PTB7-Th | FPDI-F | 1.72 | 0.92 | 0.8 | 0.4 | 8.7 | 3.2 | -0.49 | 9.00E-06 | 2.40E-01 | |
| PTB7-Th | FPDI-T | 1.72 | 0.93 | 0.79 | 0.58 | 11.95 | 6.48 | 0.12 | 9.00E-06 | 2.66E+00 | |

| Donor | Acceptor | $E_{opt}$ (eV) | $V_{oc}$ (V) | $V_{oc}$ loss (V) | FF (%) | $J_{sc}$ (mA/cm$^2$) | PCE (%) | $FF_n$ | Thickness (cm) | Q | Ref |
|---|---|---|---|---|---|---|---|---|---|---|---|
| PTB7-Th | FPDI-Se | 1.72 | 0.92 | 0.8 | 0.55 | 11.2 | 5.59 | 0.02 | 9.00E-06 | 1.86E+00 | |
| PDBT-T1 | TPH | 1.86 | 0.97 | 0.89 | 0.7 | 12.01 | 8.15 | 0.52 | 1.00E-05 | 1.70E+01 | 87 |
| | TPH-Se | 1.86 | 1 | 0.86 | 0.72 | 12.53 | 8.98 | 0.58 | 1.00E-05 | 2.23E+01 | |
| D4 | A2 | 1.98 | 0.69 | 1.29 | 0.43 | 5.02 | 1.48 | -0.36 | 6.00E-06 | 1.35E-01 | 88 |
| polythiophene derivative | A1 | 1.98 | 0.68 | 1.3 | 0.43 | 3.18 | 0.94 | -0.36 | 6.00E-06 | 1.11E-01 | |
| polythiophene derivative | A2' | 1.98 | 0.67 | 1.31 | 0.43 | 3.71 | 1.08 | -0.36 | 6.00E-06 | 1.25E-01 | 89 |
| polythiophene derivative | A3 | 1.98 | 0.69 | 1.29 | 0.4 | 2.8 | 0.77 | -0.47 | 6.00E-06 | 6.46E-02 | |
| PT1 | PC-PDI | 2 | 0.7 | 1.3 | 0.5 | 6.35 | 2.23 | -0.12 | | 0.00E+00 | 90 |
| PiI-2T | P(TP) | 1.67 | 1.01 | 0.66 | 0.5 | 6.91 | 3.48 | -0.16 | 1.00E-05 | 9.37E-01 | |
| PiI-2F | P(TP) | 1.62 | 0.96 | 0.66 | 0.51 | 5.59 | 2.71 | -0.12 | 1.00E-05 | 1.07E+00 | |
| PiI-tT | P(TP) | 1.64 | 0.97 | 0.67 | 0.42 | 4.1 | 1.67 | -0.43 | 1.80E-05 | 2.22E+00 | 91 |
| PiI-BDT | P(TP) | 1.74 | 1.07 | 0.67 | 0.45 | 6.54 | 3.12 | -0.33 | 1.20E-05 | 8.14E-01 | |
| PiI-2T-PSS | | 1.67 | 1.04 | 0.63 | 0.46 | 8.77 | 4.21 | -0.30 | 9.30E-06 | 4.69E-01 | |
| PTB7-Th | PDI-V | 1.66 | 0.74 | 0.92 | 0.63 | 15.8 | 7.3 | 0.33 | 1.20E-05 | 2.86E+01 | 92 |
| PTzBI | N2200 | 1.59 | 0.85 | 0.74 | 0.7 | 14.7 | 9.16 | 0.55 | 1.03E-05 | 3.09E+01 | 93 |
| J50 | N2200 | 1.55 | 0.6 | 0.95 | 0.59 | 13.93 | 4.8 | 0.24 | 1.20E-05 | 2.85E+01 | 94 |
| J51 | N2200 | 1.55 | 0.83 | 0.72 | 0.7 | 14.18 | 8.1 | 0.55 | 1.20E-05 | 5.57E+01 | |
| PBDTBDD | N2200 | 1.69 | 0.82 | 0.87 | 0.43 | 6.8 | 2.3 | -0.38 | 9.00E-06 | 4.30E-01 | 95 |
| PBDTBDD-T | N2200 | 1.6 | 0.87 | 0.73 | 0.575 | 11.7 | 5.6 | 0.11 | 9.00E-06 | 2.93E+00 | |
| P3HT | N2200 | 1.62 | 0.53 | 1.09 | 0.51 | 0.67 | 0.18 | -0.03 | 7.50E-06 | 6.06E-01 | 96 |
| PTB7-Th | P(NDI2OD−T) | 1.63 | 0.8 | 0.83 | 0.43 | 8.85 | 2.87 | -0.38 | 1.10E-05 | 1.05E+00 | |
| | P(NDI2OD−T2)[N2200] | 1.6 | 0.79 | 0.81 | 0.37 | 10.99 | 3.18 | -0.58 | 1.10E-05 | 4.70E-01 | 97 |
| | P(NDI2OD−TVT) | 1.58 | 0.84 | 0.74 | 0.43 | 11.4 | 4.13 | -0.38 | 1.10E-05 | 1.06E+00 | |
| PTB7-Th | N2200 | 1.63 | 0.81 | 0.82 | 0.49 | 8.7 | 3.4 | -0.17 | 9.50E-06 | 1.32E+00 | |
| | PNDI-T10 | 1.63 | 0.83 | 0.8 | 0.71 | 12.5 | 7.4 | 0.59 | 9.50E-06 | 2.75E+01 | 98 |
| | PNDI-T20 | 1.63 | 0.83 | 0.8 | 0.52 | 9.2 | 4.1 | -0.07 | 9.50E-06 | 1.83E+00 | |
| | PNDI-T50 | 1.64 | 0.83 | 0.81 | 0.48 | 5.2 | 2.1 | -0.21 | 9.50E-06 | 8.48E-01 | |
| PBDTT-TT-F | P(NDI2OD-T2) | 1.55 | 0.79 | 0.76 | 0.56 | 11.92 | 5.21 | 0.08 | 1.10E-05 | 6.36E+00 | 99 |
| PBDTT-TT-F | P(NDI2OD-FT2) | 1.63 | 0.81 | 0.82 | 0.63 | 12.32 | 6.08 | 0.31 | 1.10E-05 | 1.45E+01 | |

| Donor | Acceptor | $E_{opt}$ (eV) | $V_{oc}$ (V) | $V_{oc}$ loss (V) | FF (%) | $J_{sc}$ (mA/cm²) | PCE (%) | FFn | Thickness (cm) | Q | Ref |
|---|---|---|---|---|---|---|---|---|---|---|---|
| | P(NDI2HD-FT2) | 1.63 | 0.8 | 0.83 | 0.55 | 11.2 | 4.75 | 0.04 | 1.10E-05 | 5.30E+00 | |
| | P(NDI2DT-FT2) | 1.63 | 0.81 | 0.82 | 0.62 | 13.53 | 6.58 | 0.28 | 1.10E-05 | 1.34E+01 | |
| PSEHTT | PNDIT | 1.88 | 0.61 | 1.27 | 0.56 | 3.8 | 1.2 | 0.12 | 7.00E-06 | 1.46E+00 | 100 |
| | PNDIS | 1.82 | 0.75 | 1.07 | 0.6 | 6.53 | 2.84 | 0.23 | 7.00E-06 | 1.82E+00 | |
| | PNDIS-HD | 1.82 | 0.76 | 1.06 | 0.55 | 7.78 | 3.16 | 0.05 | 7.00E-06 | 1.03E+00 | |
| PBDTT-FTTE | PNDIS-HD | 1.67 | 0.8 | 0.87 | 0.48 | 18.6 | 7.2 | -0.20 | 1.15E-05 | 3.42E+00 | 101 |
| P3HT | PNDIBS | 1.54 | 0.53 | 1.01 | 0.44 | 3.79 | 0.84 | -0.29 | 9.00E-06 | 1.08E+00 | 102 |
| PTB7-Th | PNDIT-HD | 1.64 | 0.79 | 0.85 | 0.56 | 13.46 | 5.96 | 0.08 | 1.00E-05 | 4.84E+00 | 103 |
| | PNDIT-OD | 1.64 | 0.8 | 0.84 | 0.53 | 11.97 | 5.05 | -0.03 | 1.00E-05 | 3.09E+00 | |
| | PNDIT-DT | 1.64 | 0.81 | 0.83 | 0.52 | 7.81 | 3.25 | -0.07 | 1.00E-05 | 2.15E+00 | |
| PBDTTT-CT | 10PDI | 1.65 | 0.77 | 0.88 | 0.51 | 6.55 | 2.8 | -0.09 | 1.50E-05 | 8.16E+00 | 104 |
| | 30PDI | 1.65 | 0.76 | 0.89 | 0.42 | 15.26 | 5.1 | -0.41 | 1.50E-05 | 4.00E+00 | |
| | 50PDI | 1.65 | 0.73 | 0.92 | 0.38 | 9.44 | 2.58 | -0.54 | 1.25E-05 | 9.77E-01 | |
| PTB7-Th | IC-2IDT-IC | 1.63 | 0.93 | 0.7 | 0.37 | 12.49 | 4.26 | -0.59 | 1.10E-05 | 3.44E-01 | |
| | IC-3IDT-IC | 1.63 | 0.93 | 0.7 | 0.32 | 3.31 | 1.01 | -0.76 | 1.10E-05 | 5.87E-02 | |
| PDBT-T1 | IC-1IDT-IC | 1.71 | 0.92 | 0.79 | 0.6 | 12.84 | 7.02 | 0.19 | 9.00E-06 | 3.60E+00 | 105 |
| | IC-2IDT-IC | 1.63 | 1.02 | 0.61 | 0.48 | 5.13 | 2.42 | -0.23 | 9.00E-06 | 4.29E-01 | |
| | IC-C6IDT-IC | 1.64 | 0.85 | 0.79 | 0.68 | 15.2 | 8.95 | 0.48 | 1.05E-05 | 2.43E+01 | |
| PTB7-Th | IHIC | 1.44 | 0.75 | 0.69 | 0.67 | 18.75 | 9.41 | 0.47 | 1.00E-05 | 2.83E+01 | 106 |
| PTB7-Th | ITIC | 1.64 | 0.81 | 0.83 | 0.59 | 14.21 | 6.58 | 0.18 | 1.00E-05 | 6.72E+00 | 107 |
| J71 | ITIC | 1.66 | 0.94 | 0.72 | 0.68 | 17.4 | 11.2 | 0.46 | 1.00E-05 | 1.65E+01 | 108 |
| PBDB-TS1 | ITIC | 1.65 | 0.9 | 0.75 | 0.71 | 16.7 | 10.68 | 0.57 | 1.00E-05 | 2.96E+01 | 7 |
| J61 | ITIC | 1.65 | 0.9 | 0.75 | 0.64 | 17.72 | 10.28 | 0.33 | 1.20E-05 | 2.03E+01 | 109 |
| | ITIC-m | 1.65 | 0.91 | 0.74 | 0.69 | 18.31 | 11.49 | 0.50 | 1.20E-05 | 4.09E+01 | |
| J51 | ITIC | 1.64 | 0.81 | 0.83 | 0.68 | 16.33 | 9.07 | 0.49 | | | 110 |
| J50 | ITIC | 1.65 | 0.71 | 0.94 | 0.51 | 12.55 | 4.61 | -0.08 | 1.40E-05 | 1.09E+01 | |
| J52 | ITIC | 1.65 | 0.73 | 0.92 | 0.58 | 12.45 | 5.26 | 0.16 | | | 111 |
| J60 | ITIC | 1.65 | 0.92 | 0.73 | 0.59 | 15.84 | 8.67 | 0.16 | | | |
| PBDB-TS1 | NFBDT | 1.6 | 0.87 | 0.73 | 0.67 | 17.85 | 10.42 | 0.44 | 1.00E-05 | 1.79E+01 | 112 |
| FTAZ | INIC | 1.63 | 0.96 | 0.67 | 0.58 | 13.51 | 7.5 | 0.12 | 1.00E-05 | 3.78E+00 | 113 |
| FTAZ | INIC1 | 1.61 | 0.93 | 0.68 | 0.64 | 16.63 | 9.9 | 0.32 | 1.00E-05 | 9.53E+00 | |

| Donor | Acceptor | $E_{opt}$ (eV) | $V_{oc}$ (V) | $V_{oc}$ loss (V) | FF (%) | $J_{sc}$ (mA/cm²) | PCE (%) | FFn | Thickness (cm) | Q | Ref |
|---|---|---|---|---|---|---|---|---|---|---|---|
| FTAZ | INIC2 | 1.58 | 0.9 | 0.68 | 0.67 | 17.56 | 10.6 | 0.43 | 1.00E-05 | 1.62E+01 | |
| FTAZ | INIC3 | 1.54 | 0.86 | 0.68 | 0.67 | 19.44 | 11.2 | 0.44 | 1.00E-05 | 1.93E+01 | |
| PTB7-Th | IDT-IC | 1.67 | 0.83 | 0.84 | 0.4 | 9.53 | 3.05 | -0.48 | | | 114 |
| | IDTIDT-IC | 1.6 | 0.94 | 0.66 | 0.47 | 14.49 | 6.25 | -0.25 | 1.00E-05 | 1.11E+00 | |
| J51 | IDTIDSe-IC | 1.6 | 0.91 | 0.69 | 0.58 | 15.16 | 7.84 | 0.12 | 1.08E-05 | 5.98E+00 | 115 |
| P3HT | IDT-2BR | 1.76 | 0.84 | 0.92 | 0.68 | 8.91 | 5.04 | 0.48 | 8.00E-06 | 7.45E+00 | 116 |
| PTB7-Th | IDT-2BR | 1.65 | 1.03 | 0.62 | 0.65 | 14.5 | 9.7 | 0.34 | 1.00E-05 | 7.76E+00 | 117 |
| P3HT | BTA1 | 1.65 | 1.02 | 0.63 | 0.7 | 7.34 | 5.24 | 0.51 | | | 118 |
| PTB7-Th | ATT-1 | 1.59 | 0.87 | 0.72 | 0.7 | 16.48 | 9.89 | 0.54 | 1.30E-05 | 6.88E+01 | 119 |
| PBDTTT-C-T | IDT–2BM | 1.64 | 0.76 | 0.88 | 0.54 | 9.81 | 4.19 | 0.01 | | | 120 |
| | IDTT–2BM | 1.62 | 0.85 | 0.77 | 0.57 | 9.67 | 4.7 | 0.10 | | | |
| P3HT | IDT-2PDI | 1.62 | 0.7 | 0.92 | 0.67 | 5.58 | 2.61 | 0.49 | | | 121 |
| BDT-2DPP | IDT-2PDI | 1.7 | 0.95 | 0.75 | 0.42 | 7.75 | 3.12 | -0.42 | | | |
| PBDB-T | IT-M | 1.66 | 0.94 | 0.72 | 0.73 | 16.75 | 11.48 | 0.63 | 1.00E-05 | 3.74E+01 | 122 |
| | IT-DM | 1.69 | 0.97 | 0.72 | 0.71 | 15.82 | 10.79 | 0.55 | 1.00E-05 | 2.30E+01 | |
| PBDB-T-SF | IT-4F | 1.57 | 0.86 | 0.71 | 0.71 | 20.88 | 13.1 | 0.58 | 1.00E-05 | 3.81E+01 | 123 |
| PBDB-T-2Cl | IT-4F | 1.57 | 0.86 | 0.71 | 0.77 | 21.8 | 14 | 0.78 | 1.00E-05 | 1.53E+02 | 124 |
| PBDB-T-2F(PM6) | IT-4F | 1.57 | 0.84 | 0.73 | 0.76 | 20.81 | 13.1 | 0.76 | 1.00E-05 | 1.23E+02 | |
| PBDB-T-2Cl | ITC-2Cl | 1.54 | 0.92 | 0.62 | 0.74 | 19.64 | 13.36 | 0.67 | 1.00E-05 | 5.31E+01 | 125 |
| PBDB-T-2Cl | IXIC-4Cl | 1.31 | 0.8 | 0.51 | 0.7 | 21.19 | 11.65 | 0.56 | 1.00E-05 | 4.03E+01 | |
| PBDB-T | ITCC | 1.76 | 1.01 | 0.75 | 0.71 | 15.2 | 11 | 0.55 | 1.00E-05 | 2.01E+01 | 126 |
| PBT1-EH | ITIC | 1.65 | 0.99 | 0.66 | 0.63 | 15.2 | 9.8 | 0.28 | 1.10E-05 | 9.50E+00 | 127 |
| | ITCPTC | 1.63 | 0.95 | 0.68 | 0.75 | 16.2 | 11.4 | 0.69 | 1.10E-05 | 7.52E+01 | |
| J61 | m-ITIC | 1.65 | 0.9 | 0.75 | 0.7 | 18.3 | 11.49 | 0.53 | 1.20E-05 | 4.95E+01 | 109 |
| PTB7-Th | ITIC-Th | 1.65 | 0.8 | 0.85 | 0.68 | 15.95 | 8.5 | 0.49 | 1.20E-05 | 4.76E+01 | 128 |
| PDBT-T1 | ITIC-Th | 1.65 | 0.88 | 0.77 | 0.67 | 16.17 | 9.3 | 0.44 | 1.00E-05 | 1.65E+01 | |
| FTAZ | ITIC-Th | 1.68 | 0.91 | 0.77 | 0.61 | 15.67 | 8.67 | 0.23 | 1.20E-05 | 1.27E+01 | 129 |
| | ITIC-Th1 | 1.59 | 0.85 | 0.74 | 0.73 | 19.2 | 11.9 | 0.65 | 1.20E-05 | 1.05E+02 | |
| FTAZ | ITIC1 | 1.61 | 0.92 | 0.69 | 0.56 | 15.67 | 8.09 | 0.05 | 9.00E-06 | 2.47E+00 | 130 |
| | ITIC2 | 1.59 | 0.92 | 0.67 | 0.62 | 18.63 | 10.6 | 0.26 | 9.00E-06 | 5.55E+00 | |
| PTB7-Th | IDIC | 1.63 | 0.81 | 0.82 | 0.56 | 10.78 | 5.1 | 0.07 | 1.00E-05 | 4.07E+00 | 131 |

| Donor | Acceptor | $E_{opt}$ (eV) | $V_{oc}$ (V) | $V_{oc}$ loss (V) | FF (%) | $J_{sc}$ (mA/cm$^2$) | PCE (%) | FFn | Thickness (cm) | Q | Ref |
|---|---|---|---|---|---|---|---|---|---|---|---|
| PDCBT | | 1.65 | 0.81 | 0.84 | 0.64 | 11.12 | 6.13 | 0.35 | 1.00E-05 | 1.13E+01 | |
| J51 | | 1.65 | 0.8 | 0.85 | 0.66 | 12.08 | 6.75 | 0.42 | 1.00E-05 | 1.62E+01 | |
| PDBT-T1 | | 1.61 | 0.83 | 0.78 | 0.73 | 16.68 | 10.16 | 0.66 | 1.00E-05 | 5.60E+01 | |
| PTFBDT-BZS | | 1.62 | 0.91 | 0.71 | 0.71 | 17.16 | 10.97 | 0.57 | 1.00E-05 | 2.90E+01 | |
| PTB7-Th | IDT-2BR1 | 1.63 | 0.95 | 0.68 | 0.6 | 15.2 | 8.6 | 0.19 | 8.00E-06 | 2.39E+00 | 132 |
| | IDT-2BR | 1.64 | 0.99 | 0.65 | 0.6 | 13 | 7.6 | 0.18 | 8.00E-06 | 1.99E+00 | |
| P3HT | O-IDTBR | 1.58 | 0.72 | 0.86 | 0.6 | 13.9 | 6.3 | 0.23 | 7.50E-06 | 3.77E+00 | 133 |
| | EH-IDTBR | 1.71 | 0.76 | 0.95 | 0.62 | 12.1 | 6 | 0.29 | 7.50E-06 | 3.93E+00 | |
| PTB7-Th | IEIC1 | 1.6 | 0.84 | 0.76 | 0.44 | 9.9 | 3.7 | -0.35 | | | 134 |
| | IEIC2 | 1.6 | 0.91 | 0.69 | 0.5 | 11.1 | 5.4 | -0.15 | | | |
| | IEIC3 | 1.59 | 0.93 | 0.66 | 0.58 | 12.8 | 7.4 | 0.12 | | | |
| | IEIC | 1.6 | 0.95 | 0.65 | 0.49 | 12.3 | 6 | -0.19 | | | |
| PBDTTT-E-T | IEICO | 1.43 | 0.82 | 0.61 | 0.58 | 17.7 | 8.3 | 0.14 | 9.50E-06 | 5.37E+00 | 135 |
| | IEIC | 1.6 | 0.9 | 0.7 | 0.47 | 11.7 | 4.7 | -0.25 | 1.05E-05 | 1.31E+00 | |
| PBDTTT-EFT | IEICO-4F | 1.34 | 0.74 | 0.6 | 0.59 | 22.4 | 9.7 | 0.19 | 1.00E-05 | 1.07E+01 | 136 |
| J52 | IEICO-4F | 1.34 | 0.73 | 0.61 | 0.59 | 21.4 | 9.3 | 0.20 | 1.00E-05 | 1.09E+01 | |
| PTB7-Th | COi8DFIC | 1.26 | 0.7 | 0.56 | 0.71 | 27.3 | 13.4 | 0.63 | 1.00E-05 | 8.58E+01 | 137 |
| PTB7-Th | CTIC-4F | 1.37 | 0.7 | 0.67 | 0.64 | 23.4 | 10 | 0.38 | | | 138 |
| PTB7-Th | CO1-4F | 1.27 | 0.64 | 0.63 | 0.64 | 24.8 | 10 | 0.40 | | | |
| PTB7-Th | COTIC-4F | 1.15 | 0.57 | 0.58 | 0.61 | 20.7 | 6.9 | 0.32 | | | |
| PBDB-TF | IEICO-4F | 1.35 | 0.84 | 0.51 | 0.51 | 5.62 | 2.66 | -0.10 | | | 139 |
| PM7 | IDT6CN-M | 1.68 | 1.05 | 0.63 | 0.77 | 16.4 | 12.4 | 0.74 | | | 140 |
| PM6 | TOBDT | 1.49 | 0.88 | 0.61 | 0.67 | 17.8 | 11 | 0.44 | | | 141 |
| PBDB-T | BCPT-4F | 1.39 | 0.78 | 0.61 | 0.7 | 22.3 | 12.2 | 0.57 | | | 142 |
| PBDB-T | IDT2-DFIC | 1.54 | 0.91 | 0.63 | 0.69 | 15.5 | 10 | 0.50 | | | 143 |
| PM6 | Y6 (BTP-4F) | 1.41 | 0.83 | 0.58 | 0.75 | 25.3 | 15.6 | 0.72 | 1.50E-05 | 4.50E+02 | 6 |
| PBDB-T | Y1 | 1.43 | 0.88 | 0.55 | 0.7 | 21.4 | 13 | 0.54 | 1.08E-05 | 3.96E+01 | 144 |
| | Y1-4F | 1.38 | 0.74 | 0.64 | 0.57 | 22.7 | 9.6 | 0.12 | 0.0000108 | 1.10E+01 | |
| PM6 | Y1 | 1.43 | 0.92 | 0.51 | 0.55 | 12.9 | 6.6 | 0.02 | 0.0000108 | 3.77E+00 | |
| | Y1-4F | 1.38 | 0.83 | 0.55 | 0.68 | 24.81 | 14.4 | 0.48 | 0.0000108 | 3.68E+01 | |
| PTQ7 | Y6 | 1.4 | 0.71 | 0.69 | 0.43 | 18.65 | 5.69 | -0.36 | 1.30E-05 | 3.63E+00 | 145 |

| Donor | Acceptor | $E_{opt}$ (eV) | $V_{oc}$ (V) | $V_{oc}$ loss (V) | FF (%) | $J_{sc}$ (mA/cm²) | PCE (%) | $FF_n$ | Thickness (cm) | Q | Ref |
|---|---|---|---|---|---|---|---|---|---|---|---|
| PTQ8 | Y6 | 1.4 | 0.89 | 0.51 | 0.32 | 3.11 | 0.86 | -0.76 | 1.30E-05 | 1.13E-01 | |
| PTQ9 | Y6 | 1.4 | 0.82 | 0.58 | 0.54 | 23.72 | 10.3 | 0.00 | 1.30E-05 | 1.15E+01 | |
| PTQ10 | Y6 | 1.4 | 0.87 | 0.53 | 0.75 | 24.81 | 15.97 | 0.71 | 1.30E-05 | 2.27E+02 | |
| PM6 | BTP-4Cl | 1.4 | 0.87 | 0.53 | 0.74 | 25.2 | 16.1 | 0.68 | 1.10E-05 | 1.01E+02 | 146 |
| PM6 | N-C11 | 1.45 | 0.85 | 0.6 | 0.71 | 21.18 | 12.47 | 0.58 | 1.05E-05 | 4.72E+01 | 147 |
| | N3 | 1.39 | 0.84 | 0.55 | 0.74 | 25.64 | 15.79 | 0.69 | 1.05E-05 | 9.82E+01 | |
| | N4 | 1.42 | 0.82 | 0.6 | 0.7 | 24.61 | 13.63 | 0.55 | 1.05E-05 | 4.77E+01 | |
| Pt0 | Y6 | 1.37 | 0.8 | 0.57 | 0.65 | 24.63 | 12.6 | 0.39 | 1.00E-05 | 2.00E+01 | 148 |
| Pt5 | Y6 | 1.37 | 0.8 | 0.57 | 0.73 | 25.37 | 14.66 | 0.66 | 1.00E-05 | 7.85E+01 | |
| Pt10 | Y6 | 1.37 | 0.81 | 0.56 | 0.76 | 25.98 | 16.02 | 0.76 | 1.00E-05 | 1.59E+02 | |
| Pt15 | Y6 | 1.37 | 0.82 | 0.55 | 0.74 | 25.56 | 15.41 | 0.69 | 1.00E-05 | 9.00E+01 | |
| PBDB-TF(PM6) | BTP-4F-12 | 1.38 | 0.85 | 0.53 | 0.74 | 25 | 15.9 | 0.68 | 1.00E-05 | 7.84E+01 | 149 |
| S1 | Y6 | 1.41 | 0.88 | 0.53 | 0.74 | 24.9 | 15.8 | 0.68 | 1.05E-05 | 8.23E+01 | 150 |
| S4 | Y6 | 1.41 | 0.93 | 0.48 | 0.49 | 12.5 | 5.6 | -0.18 | 1.05E-05 | 1.61E+00 | |
| ES1 | Y6 | 1.41 | 0.87 | 0.54 | 0.71 | 24.3 | 14.8 | 0.58 | 1.05E-05 | 4.70E+01 | |
| PE61 | Y6 | 1.39 | 0.66 | 0.73 | 0.55 | 23.4 | 8.4 | 0.07 | | | 151 |
| PE62 | Y6 | 1.39 | 0.78 | 0.61 | 0.62 | 24.6 | 11.8 | 0.29 | | | |
| PE63 | Y6 | 1.39 | 0.83 | 0.56 | 0.64 | 24.7 | 13 | 0.34 | | | |
| PBDB-TF(PM6) | AQx-1 | 1.41 | 0.89 | 0.52 | 0.67 | 21.9 | 13 | 0.43 | 1.10E-05 | 2.60E+01 | 152 |
| | AQx-2 | 1.4 | 0.86 | 0.54 | 0.76 | 25.2 | 16.4 | 0.75 | 1.10E-05 | 1.72E+02 | |
| PM6 | Y11 | 1.36 | 0.85 | 0.51 | 0.72 | 26.4 | 16.5 | 0.62 | 1.30E-05 | 1.34E+02 | 31 |
| PTQ10 | DF-PCIC | 1.66 | 1.04 | 0.62 | 0.54 | 6.5 | 3.5 | -0.03 | | | 140 |
| | HC-PCIC | 1.58 | 0.94 | 0.64 | 0.68 | 16.0 | 10.2 | 0.46 | | | |
| PBDB-T | DF-PCIC | 1.64 | 0.89 | 0.75 | 0.62 | 15.3 | 8.4 | 0.26 | | | |
| PM6 | L8-BO | 1.42 | 0.87 | 0.55 | 0.81 | 25.7 | 18.0 | | | | 153 |
| | L8-HD | 1.43 | 0.88 | 0.55 | 0.78 | 24.9 | 17.1 | | | | |
| | L8-OD | 1.42 | 0.89 | 0.53 | 0.74 | 24.6 | 15.9 | | | | |
| PM6 | Y1 | 1.44 | 0.91 | 0.53 | 0.59 | 16.1 | 8.6 | 0.16 | 1.20E-05 | 5.31E+00 | This work |
| PM6 | Y11 | 1.36 | 0.84 | 0.52 | 0.65 | 21.1 | 11.4 | 0.38 | 1.20E-05 | 6.91E+01 | This work |
| PM6 | IEICO-4F | 1.36 | 0.83 | 0.53 | 0.60 | 8.7 | 4.3 | 0.21 | 1.20E-06 | 2.55E+00 | This work |
| PTO2 | Y1 | 1.44 | 0.95 | 0.49 | 0.27 | 1.6 | 0.4 | -0.93 | 1.20E-05 | 2.78E-03 | This work |

**Supplementary Table 1 | Summarized device performance, voltage loss and the Q value.**

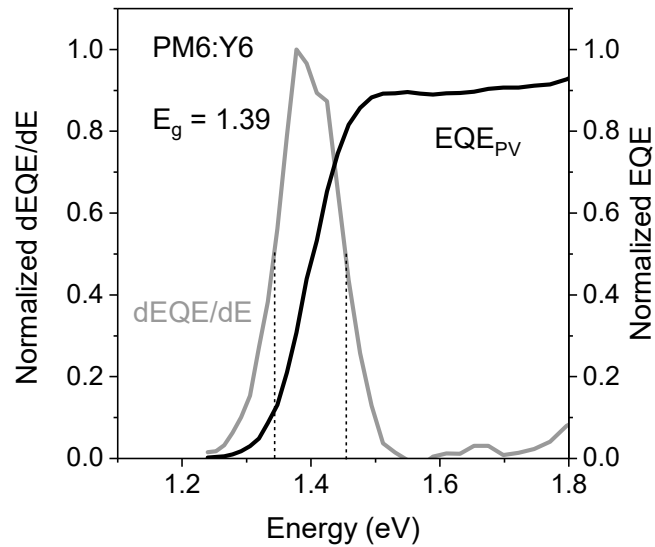

**Supplementary Figure 1 | A demo of calculating the optical gap based on the device photovoltaic external quantum efficiency (EQE$_{PV}$).** As reported in the previous literature[154], an EQE is interpreted as a superposition of a distribution of step-functions with a step at Eg having a certain probability distribution. This probability distribution can be obtained from the derivative $d$EQE/$dE$. The part where the probability is greater than half of the maximum is integrated to obtain an average gap. The EQE$_{PV}$ data of literatures were obtained by using the Digitizer tool of OriginLab.

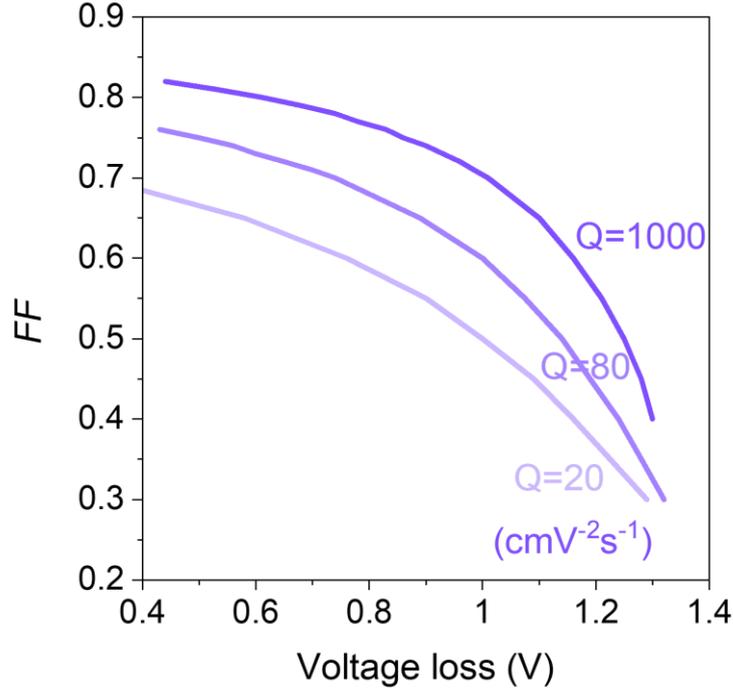

**Supplementary Figure 2 | Q value, FF and voltage loss.** A Shockley-type solar cell with an optical gap of 1.4 eV, thickness of 100 nm, and $EQE_{PV}$ of 0.8 was assumed. By setting a series of $V_{OC}$ values smaller than the SQ limit, a series of $Q$ and $FF$ values were achieved.

We make use of an approach to calculate the $Q$ value without measuring the recombination coefficient $k_2$ and the mobility $\mu$[27]. The $FF$ is normalized ($FF_n$) and mapped onto the range (-1, 1)

$$FF_n = 2\frac{FF - FF_{min}}{FF_{max} - FF_{min}} - 1 \qquad (3)$$

where $FF_{min}$ is set 0.25, corresponding to non-rectifying device with a linear JV curve, and $FF_{max}$ a simple relation proposed in an early work[21]

$$FF_{max} = \frac{aV_{OC}}{V_{OC} + b} \qquad (4)$$

where a and b are constant parameters, 0.92 and 0.09, respectively, to be found by fitting. By assuming that $FF_n$ follows hyperbolic tangent function due to the S shape of scattered data points, fitting was done with this equation

$$FF_n = \tanh\left(\alpha_{dir} \ln\left(\frac{\gamma_{dir}}{\gamma_0}\right)\right) \qquad (5)$$

where $\alpha_{dir}$ and $\gamma_0$ are fitting parameters as well. 'dir' means direct recombination or bimolecular recombination. $\gamma_{dir}$ is the collection coefficient

$$Q_{fit} = \gamma_0 \frac{d^{3.5} J_{SC}^{0.5}}{V_{OC}^2} \exp\left(\frac{\operatorname{arctanh} FF_n}{\alpha_{dir}}\right) \qquad (6)$$

where $\gamma_0$ is $2 \times 10^{17}$ $mA^{-0.5}s^{-1.2}cm^{-0.9}$, and $\alpha_{dir}$ is 0.29.

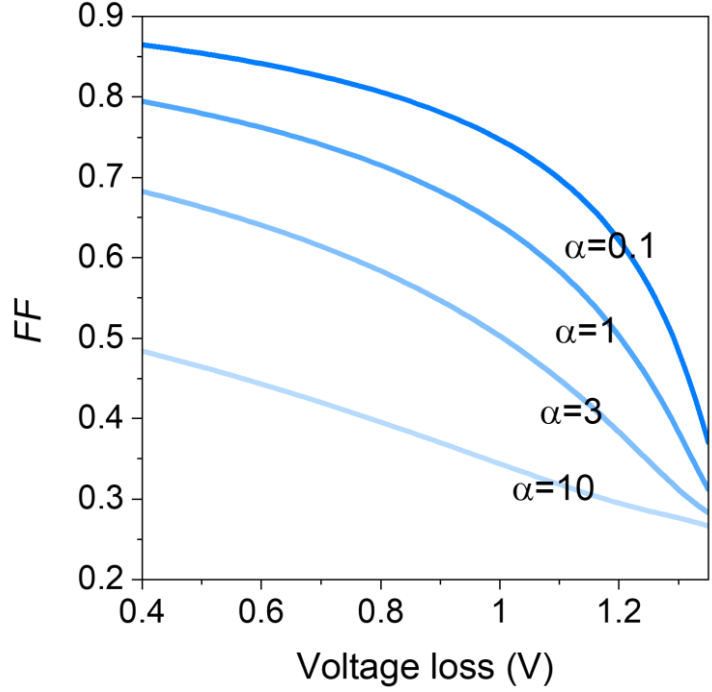

**Supplementary Figure 3 | The limit of FF versus voltage loss caused by different figure of merit α.**
A constant α allows higher FF at lower voltage loss. When α is decreased, the limit is increased overall.

The figure of merit α is defined as[29]

$$\alpha^2 = \frac{qk_2 d^3 J_G}{4\mu_{eff}^2 (K_B T)^2} \qquad (7)$$

where q is the elementary charge, and $k_B T$ is the thermal energy, $k_2$ is the bimolecular recombination coefficient, $d$ is the film thickness, $J_G$ is the generation current, $\mu_{eff}$ is the effective mobility.

With α, FF can be expressed as

$$FF = \frac{u_{OC} - \ln(0.79 + 0.66 u_{OC}^{1.2})}{u_{OC} + 1} \qquad (8)$$

$$u_{OC} = \frac{qV_{OC}}{(\alpha + 1)k_B T} \qquad (9)$$

where $u_{OC}$ is the reduced voltage.

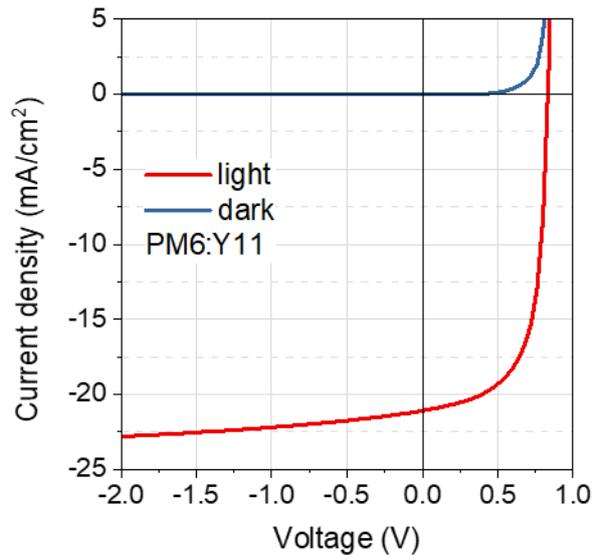
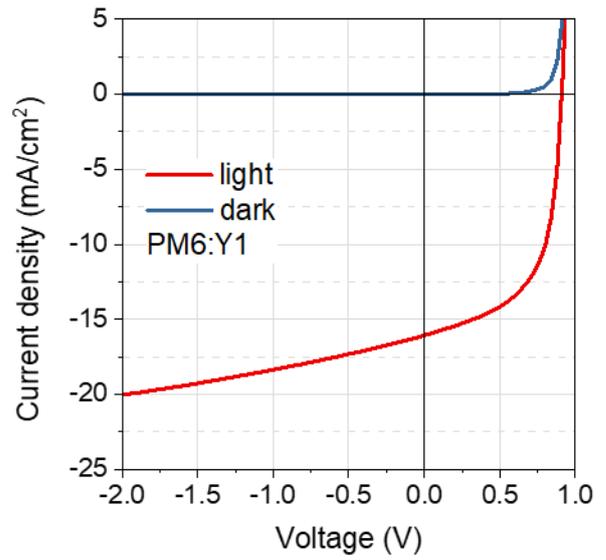
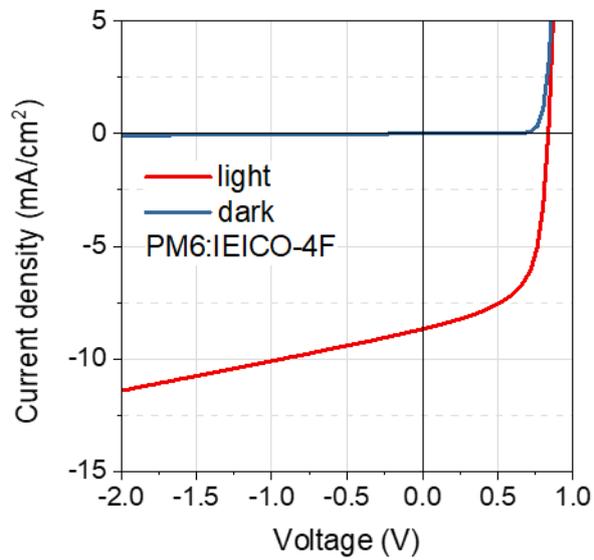
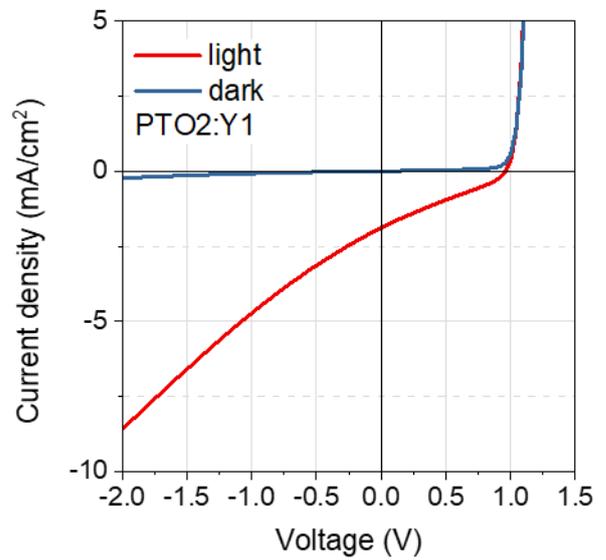

**Supplementary Figure 4 | Device performance.** PM6:Y11, PM6:Y1, PM6:IEICO-4F and PTO2:Y1 show good diode curves in dark conditions, while their performance under illuminated conditions is very different.

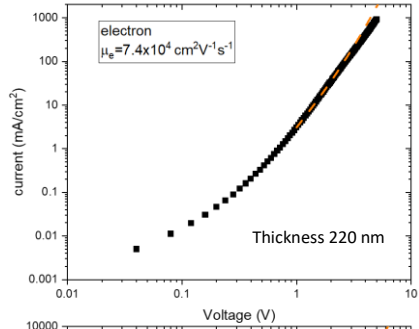 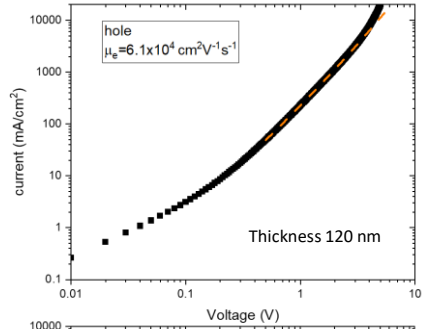

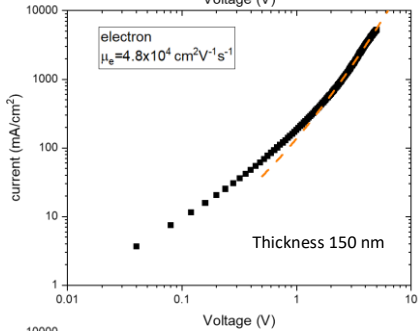 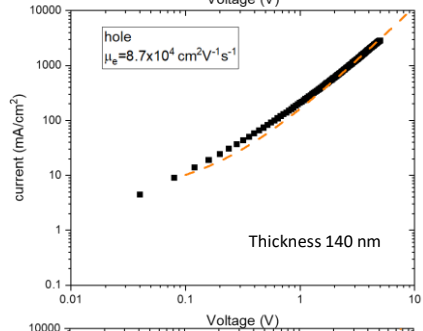

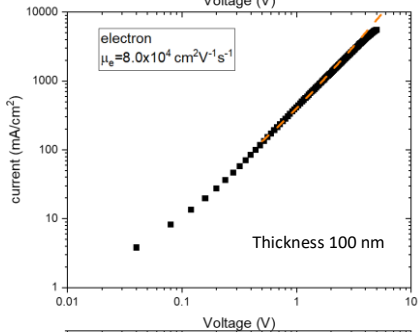 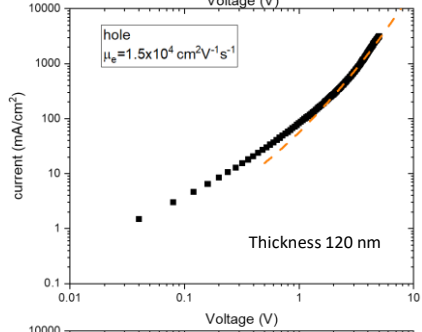

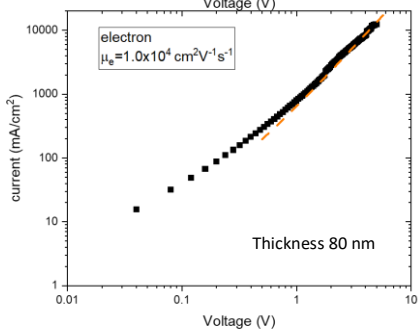 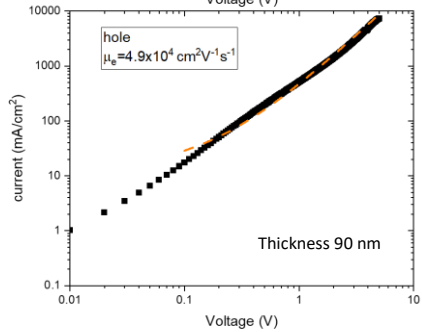

**Supplementary Figure 5 | Voltage-current data of single carrier devices.** Single-carrier devices of PM6:Y11, PM6:Y1, PM6:IEICO-4F and PTO2:Y1 have been fabricated and measured. The voltage-current data and film thickness are used for mobility evaluations. The data were modeled according to the report by Felekidis et al[155].

|  | PM6:Y11 | PM6:Y1 | PM6:IEICO-4F | PTO2:Y1 |
|---|---|---|---|---|
| $k_2$ (m$^{-3}$s$^{-1}$) | 6.63×10$^{-17}$ | 2.11×10$^{-16}$ | 2.23×10$^{-16}$ | 6.10×10$^{-16}$ |
| $d$ (m) | 1.00×10$^{-07}$ | 1.00×10$^{-07}$ | 9.00×10$^{-08}$ | 8.00×10$^{-08}$ |
| $J_{SC}$ (Am$^{-2}$) | 220 | 160 | 90 | 20 |
| $\mu_e$ (m$^2$V$^{-1}$s$^{-1}$) | 7.40×10$^{-08}$ | 4.80×10$^{-08}$ | 8.00×10$^{-08}$ | 1.00×10$^{-08}$ |
| $\mu_h$ (m$^2$V$^{-1}$s$^{-1}$) | 6.10×10$^{-08}$ | 8.70×10$^{-08}$ | 1.50×10$^{-08}$ | 4.90×10$^{-08}$ |
| $\mu$ (m$^2$V$^{-1}$s$^{-1}$) | 6.72×10$^{-08}$ | 6.46×10$^{-08}$ | 3.46×10$^{-08}$ | 2.21×10$^{-08}$ |
| $Q$ (cmV$^{-2}$s$^{-1}$) | 5612 | 3587 | 101 | 394 |
| $A$ | 2.7 | 4.3 | 5.3 | 5.4 |
| $V_{OC}$ | 0.83 | 0.91 | 0.83 | 0.95 |
| $u_{OC}$ | 8.6 | 6.6 | 5.1 | 5.7 |
| Reconstructed FF | 0.66 | 0.61 | 0.56 | 0.58 |
| Measured FF | 0.67 | 0.60 | 0.59 | 0.27 |

**Supplementary Table 2 | The reconstruction of *FF* from α and other device properties.**

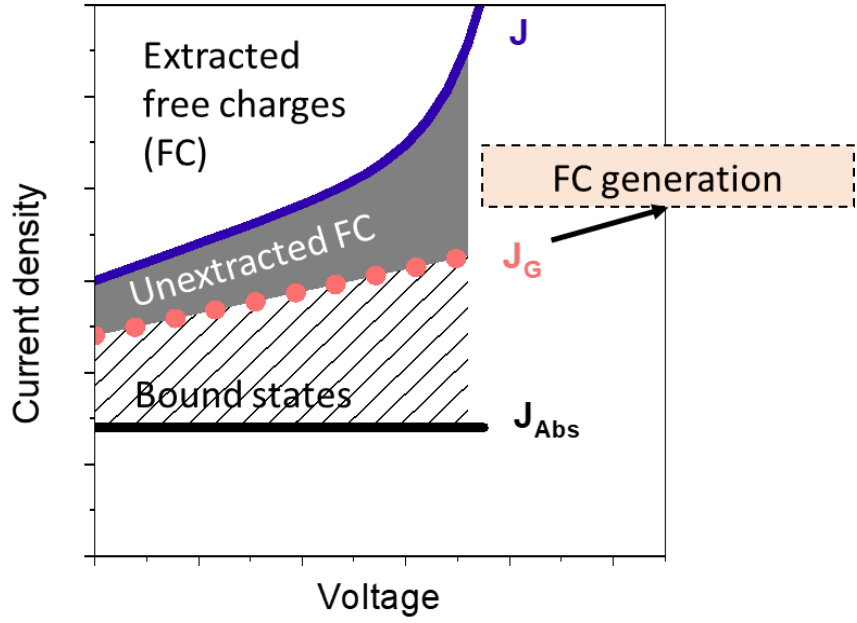

**Supplementary Figure 6 | The illustration of charge separation data.** The blue curve is the photocurrent density ($J$). The pink dots are free charge (FC) generation current density ($J_G$). The black line is the exciton generation current density ($J_{ABS}$) due to light absorption.

The total current is the sum of the injection current and the photo current.

$$J(V) = J_{\text{inj}} - J_G(V) \tag{10}$$

In the ideal case, injection current is

$$J_{\text{inj}} = J_0 \left( \exp\left(\frac{qV}{k_B T}\right) - 1 \right) \tag{11}$$

The photocurrent which is measured by TDCF depends on the generation efficiency,

$$J_G(V) = J_{\text{Abs}} \eta_{\text{int}}(V) \tag{12}$$

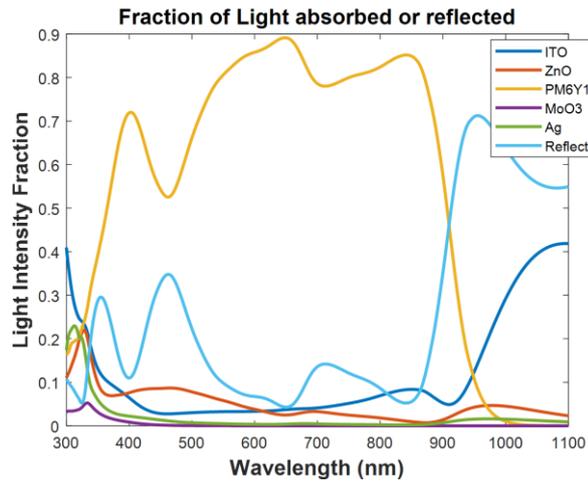 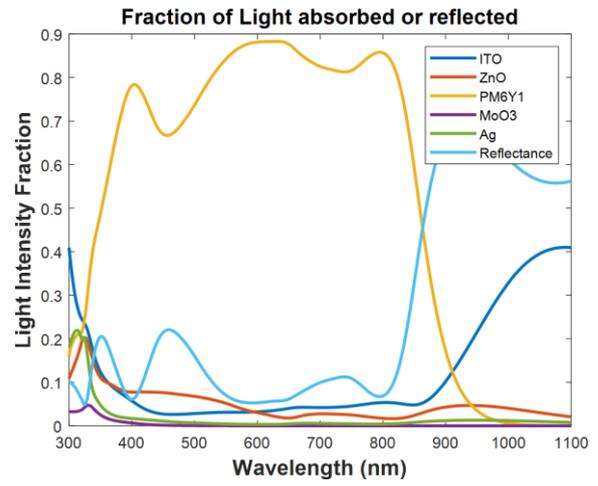 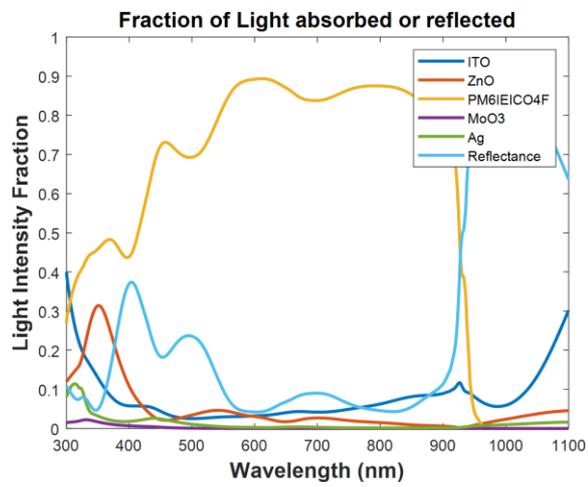 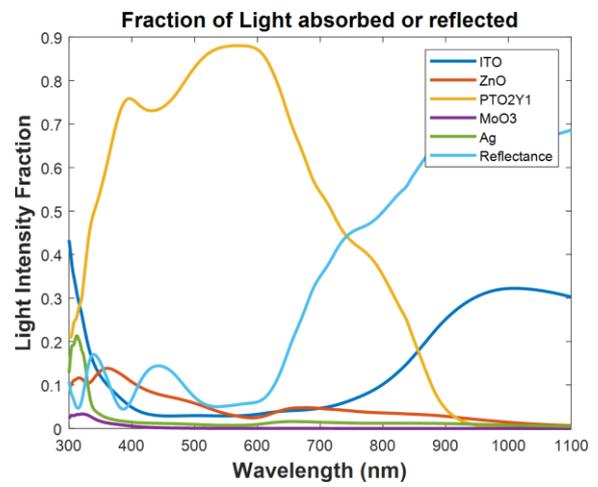

**Supplementary Figure 7 | The absorbed light of device layers.** The parameters are modeled with ellipsometry data and transfer matrix[156].

**Supplementary note 1**

We use a power series with $s = \frac{q|V-V_{OC}|/d}{q^2/(\pi\varepsilon_0\varepsilon d_{eh}^2)} = \frac{|V-V_{OC}|(\pi\varepsilon_0\varepsilon d_{eh}^2)}{qd}$, and write $J_G(V) = J_{Abs}\eta_{int}\sum_{i=0}^{\infty}\gamma_i s^i$, where $J_{Abs}$ is the generation current density due to optical absorption, $\eta_{int}$ is the internal generation efficiency on the OC condition, $\gamma_i$ represents the coefficient of order $i$ after expansion, and $d$ is the device thickness, $q$ is the elementary charge, $\varepsilon_0$ is the permittivity of free space, $\varepsilon$ is the relative permittivity and $d_{eh}$ is the electron-hole distance of the geminate pair. The equation is inspired by Braun's model[56], which has been used in the past to describe the field dependence of free charge generation from CT states in OSCs[48,157,158]. However, Braun's model has also been criticized for lack of rigor and limited scope[58]. Therefore, when it is unclear which process the electric field dependence comes from, we only use the advantage of Braun's model, that is, linear expression, and take the ratio of the external and internal electric fields as the base of the power series to perform mathematical analysis of the experimental results. When $i = 0$, the term represents the zero-field FC generation and $\gamma_0 = 1$. Compared to the *FF* losses due to the transport limit, which was discussed in zthe previous section, the terms with $i \geq 1$ are additional *FF* loss channels. Considering that $\varepsilon \approx 3$, $|V - V_{OC}| < 1$ V, $d_{eh} < 10$ nm and $d \approx 100$ nm for an OSC, we have $s \ll 1$. Also note that our experimental FC generation has linear shape in the same region. Therefore, we deduce that $J_G(V) \approx J_{Abs} \times (\eta_{int} + \eta_{int}\gamma_1 s)$ from Equation (3). Considering that there is always $\eta_{int} + \eta_{int}\gamma_1 s \leq 1$, we have $\gamma_1 s \leq \frac{1}{\eta_{int}} - 1$. Hence, we rewrite that $\gamma_1 s = \left(\frac{1}{\eta_{int}} - 1\right)\beta|V - V_{OC}|$ when $|V - V_{OC}| < 1$ and $\beta \leq 1$.

|  | PM6:Y11 | PM6:Y1 | PM6:IEICO-4F | PTO2:Y1 |
|---|---|---|---|---|
| $\alpha$ | 2.7 | 4.3 | 5.3 | 5.4 |
| $V_{OC}$ (V) | 0.83 | 0.91 | 0.83 | 0.95 |
| $\beta$ (V$^{-1}$) | 0.6 | 0.27 | 0.05 | 0.09 |
| $\eta_{int}$ | 0.72 | 0.55 | 0.34 | 0.06 |
| $J_{Abs}$ (Am$^{-2}$) | 26.7 | 25.8 | 25.1 | 20.8 |

**Supplementary Table 3 | The parameters for JV simulation.** All the parameters were deducted from experiments. Here, $\alpha$ is the FoM, $\beta$ is the defined field-dependent constant, $\eta_{int}$ is the internal generation efficiency and $J_{Abs}$ is the optical generation current density.

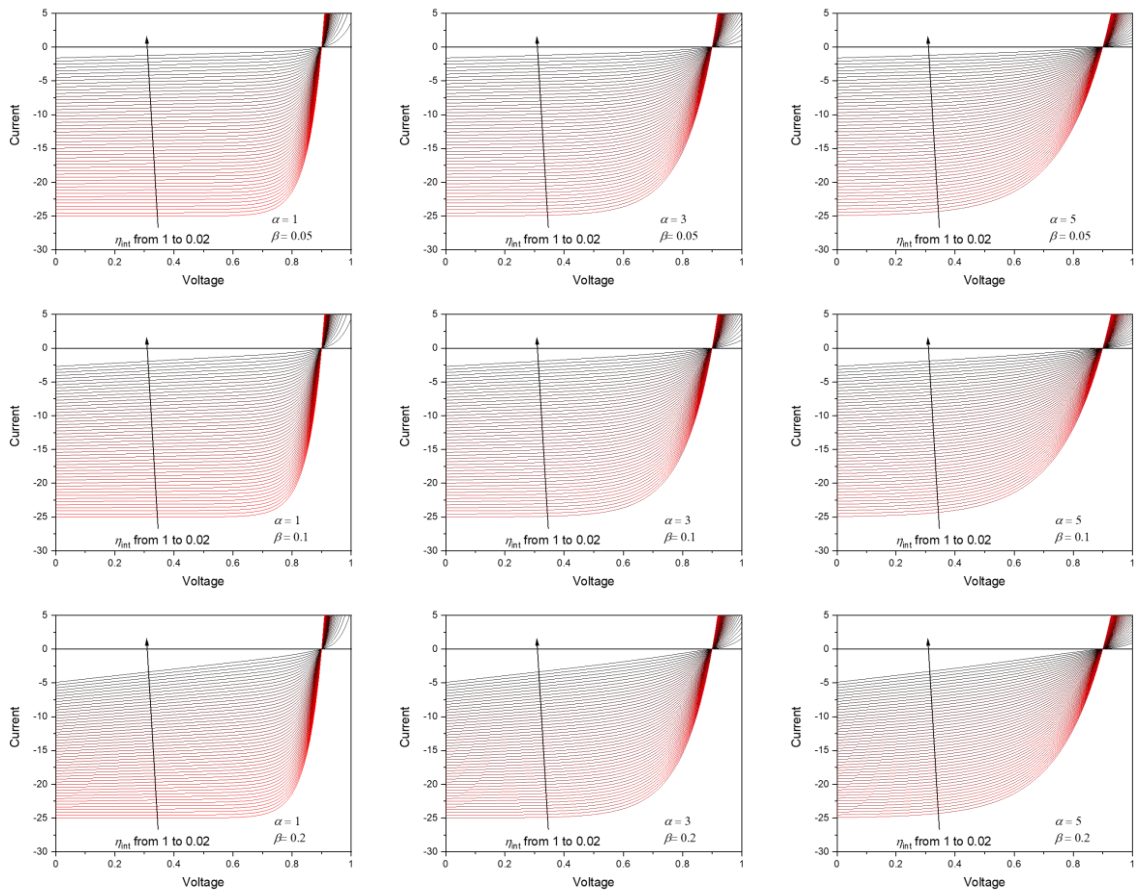

**Supplementary Figure 8 | Simulated JV curves in the presence of field-dependent geminate recombination.**

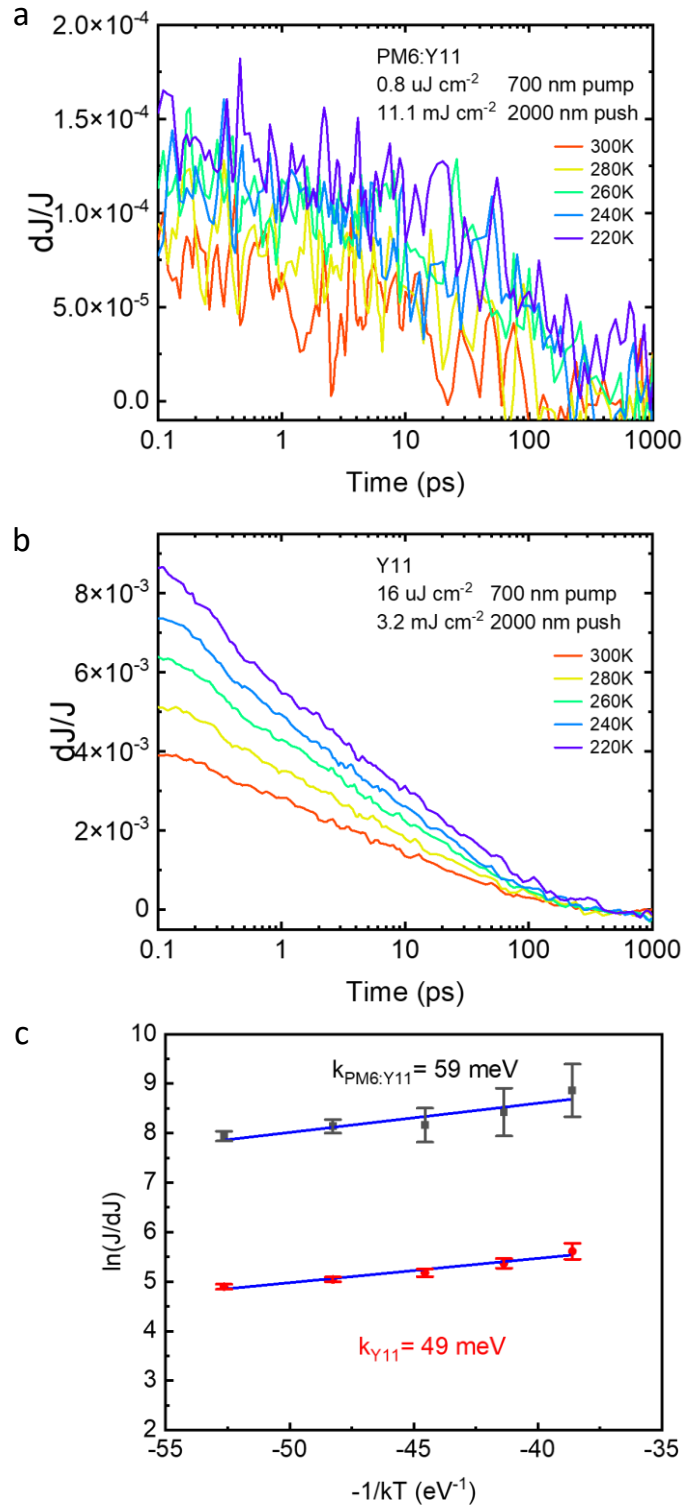

**Supplementary Figure 9 | Temperature-dependent pump-push photocurrent of blend films and neat films. a** Pump-push photocurrent of PM6:Y11. The photocurrent is the delta current when there is a 'push' pulse along with the 'pump', which means 'push' is modulated with a chopper. The time delay is between the 'push' and the 'pump'. The kinetics of blends show maximum at the beginning which indicates the geminate pairs are not generated afterwards. **b** Pump-push photocurrent of pristine Y11. **c** The temperature dependence of the pump-push photocurrent intensity integrates the whole kinetics of pump-push photocurrents. Blend and neat-acceptor devices show similar activation energy which makes sense.

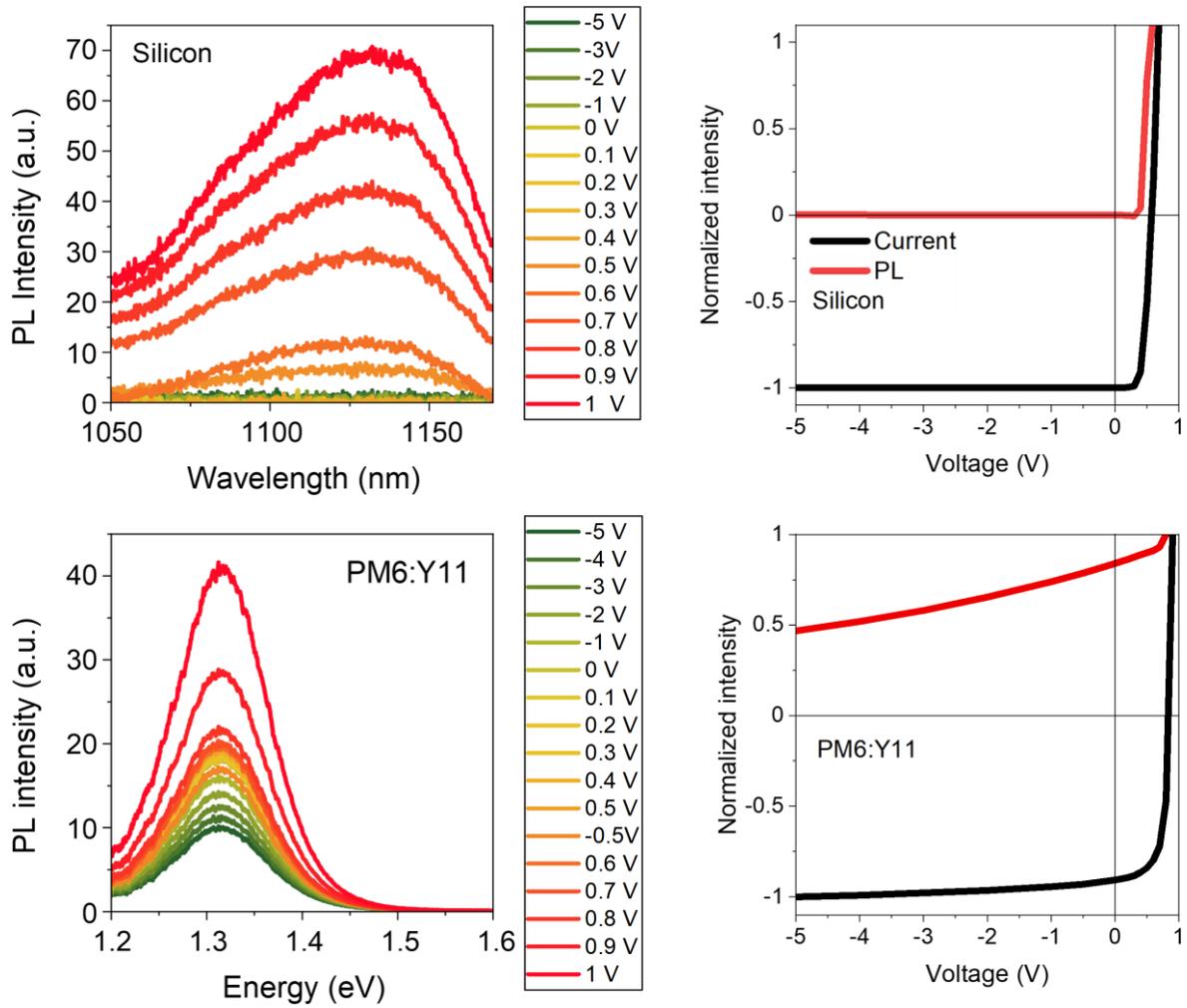

**Supplementary Figure 10 | The bias-dependent PL spectra and JV of a silicon solar cell and an OSC PM6:Y11.**

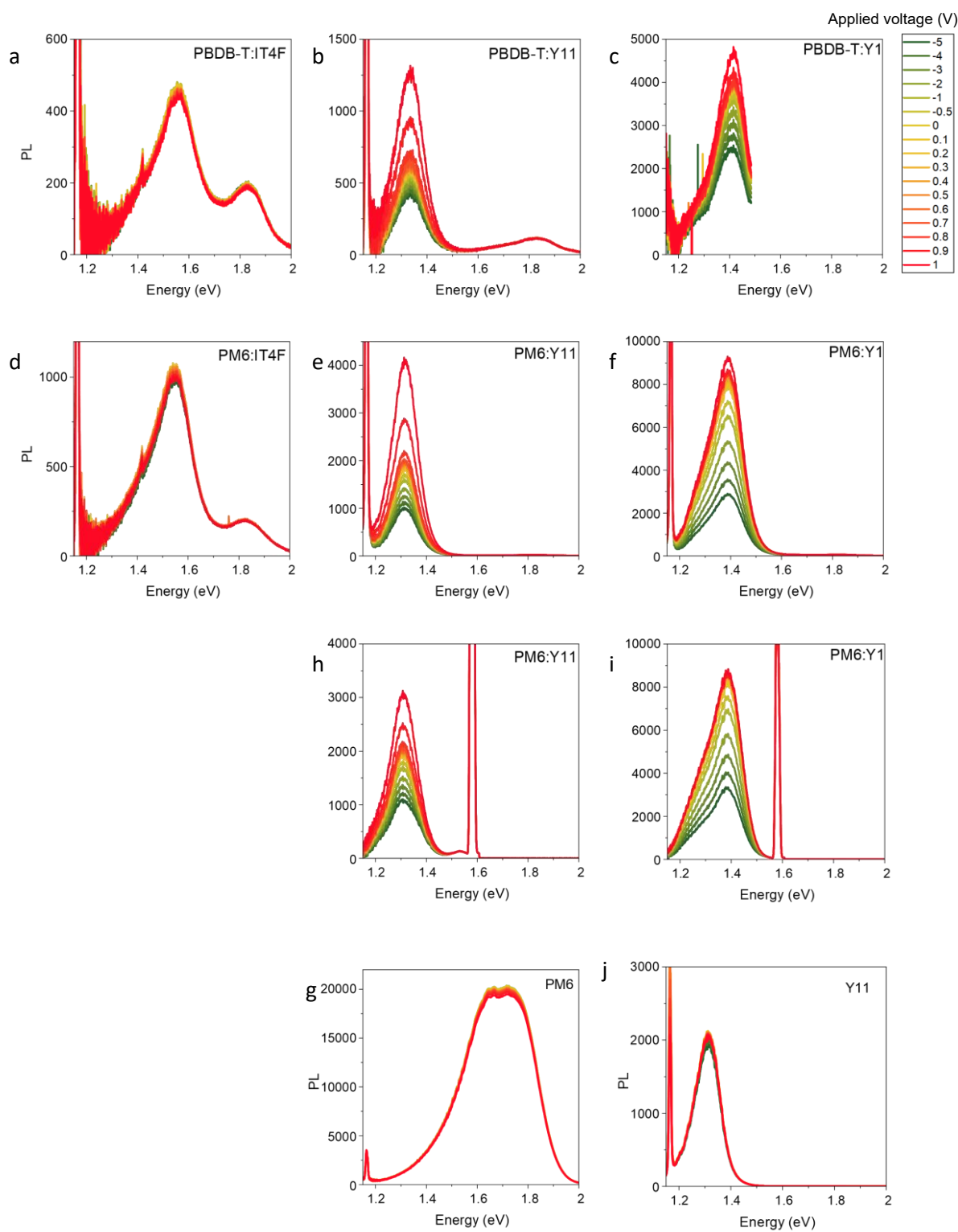

**Supplementary Figure 11 | The bias-dependent PL spectra of organic solar cells. a-f** Devices of blends are excited with a 532 nm laser. Hence, both donors and acceptors are excited. The power of the laser ensures that the devices have $J_{SC}$ equivalent to one-sun conditions. **h** and **i** Devices are excited with a 780 nm laser. Here, only acceptors are excited, which shows similar bias-dependence. **g** and **j** Devices of pristine materials, which are measured with the same parameters as the blends.

**Supplementary note 2**

Without the electric field:

$$k_{\text{Ex-CT}} = \frac{|H_{\text{Ex-CT}}|^2}{\hbar} \sqrt{\frac{\pi}{\lambda k_B T}} \exp\left(-\frac{(-\Delta E_{\text{Ex-CT}} + \lambda)^2}{4\lambda k_B T}\right) \quad (13)$$

$$k_{\text{CT-Ex}} = \frac{|H_{\text{CT-Ex}}|^2}{\hbar} \sqrt{\frac{\pi}{\lambda k_B T}} \exp\left(-\frac{(-\Delta E_{\text{CT-Ex}} + \lambda)^2}{4\lambda k_B T}\right) = k_{\text{Ex-CT}} \exp\left(\frac{\Delta E_{\text{CT-Ex}}}{k_B T}\right) \quad (14)$$

where $k_{\text{Ex-CT}}$ is the transition rate from the Ex to the CT state, $H_{\text{Ex-CT}}$ the electronic coupling from the Ex to CT state, $k_{\text{CT-Ex}}$ the back transfer rate from the CT to the Ex state, $\Delta E_{\text{Ex-CT}} = E_{\text{Ex}} - E_{\text{CT}}$ the energy difference between the Ex and CT state, $H_{\text{CT-Ex}}$ the electronic coupling from the Ex to CT state which equals $H_{\text{Ex-CT}}$, $\hbar$ the reduced Planck constant, $\lambda$ the reorganization energy, $k_B$ the Boltzmann constant, $T$ the temperature, and $\Delta E_{\text{CT-Ex}} = E_{\text{CT}} - E_{\text{Ex}}$ the energy difference between the CT and Ex state.

With the electric field:

$$\Delta E_{\text{CT}} = -\boldsymbol{\mu}_{\text{CT}} \boldsymbol{F} - \frac{1}{2}\alpha_{\text{CT}} F^2 \quad (15)$$

$$|\boldsymbol{\mu}_{\text{CT}}| = q d_{\text{CT}} \quad (16)$$

$$k_{\text{Ex-CT}} = \frac{|H_{\text{Ex-CT}}|^2}{\hbar} \sqrt{\frac{\pi}{\lambda k_B T}} \exp\left(-\frac{(-\Delta E_{\text{Ex-CT}} + \lambda + \Delta E_{\text{CT}})^2}{4\lambda k_B T}\right) \quad (17)$$

where $\boldsymbol{F}$ is the electric field, $\boldsymbol{\mu}_{\text{CT}}$ the electric dipole moment of the CT state, $\alpha_{\text{CT}}$ the polarizability of the CT state, $q$ the elementary charge, and $d_{\text{CT}}$ the distance between the positive and the negative charge. The first and second terms in eq. 13 represent the first and second order Stark effect, respectively. In eq. 15, the energy change of the singlet exciton is neglected because of the relatively small electric dipole moment and polarizability of singlet (LE) or intramolecular CT (ICT) excitons (e.g., $\mu_{\text{TCNQ,LE}} = 0$ D, $\mu_{\text{TCNQ,ICT}} = 6$ D, $\alpha_{\text{TCNQ,LE}} = 14$ Å$^3$, and $\alpha_{\text{TCNQ,ICT}} = 85$ Å$^3$)[159]. In a bulk-heterojunction blend with randomly oriented dipoles, the first order terms cancel each other, leaving only the quadratic terms. The dependence of the polarizability on the radius can be described by $\alpha_p \propto d^4$. This can be elucidated through the conventional expression for the linear polarizability, which is given by $\alpha_p \sim \frac{|\mu|^2}{\delta}$, where $\boldsymbol{\mu}$ denotes the transition dipole moment and $\delta$ represents the standard energy-level spacing of the charge carrier. For a carrier confined within a region of characteristic size $d$, the dipole moment is of the order of $qd$, and the energy-level spacing is $\sim \frac{h^2}{md^2}$, where $m$ is the mass of the charge carrier and $h$ is Planck's constant. Therefore, the resultant equation is $\alpha_p \sim d^4/a_B$, where $a_B$ denotes the Bohr radius[160]. In this work, a radius of 1.5 nm is used for the CT state according to previous simulations for PM6:Y6[161], which leads to $\alpha_{\text{CT}} = 8.5 \times 10^5$ Å$^3$ (when converted from cgs units to SI units, it needs to be multiplied by $4\pi\varepsilon_0$, where $\varepsilon_0$ is the vacuum permittivity).

The accurate values of reorganization energy $\lambda$ are currently controversial, with reported values ranging from 250 to 600 meV[37,161–163]. Inner reorganization energy of 330 meV was predicted for Y5- and Y6 aggregates with consideration of the charging of the donor upon CT formation[37]. Outer reorganization energy of 230 meV was found for PBDB-T:ITIC[163]. Note that contrastingly smaller values for the inner (161 meV) and outer (150 meV) reorganization energy were predicted for a J61:m-ITIC dimer[162], also small values of classical (100 meV) and high-frequency-mode-associated (150 meV) reorganization energy for PM6:Y6[161].

| Parameter | Value |
| --- | --- |
| $E_{CT}$ (eV) | 1.4 |
| $H_{Ex\text{-}CT}$ (eV) | 0.01 |
| $d_{CT}$ (nm) | 1.5 |
| $\alpha_{CT}$ (Å$^3$) | $8.5 \times 10^5$ |
| $\lambda$ (eV) | 0.5 |

**Supplementary note 3**

Devices are simulated with

$$\frac{\partial n(x)}{\partial t} = k_{\text{dis}} \times [\xi(x) - \xi_0] - k_{\text{rec}} \times [n(x) - n_0] \times [p(x) - p_0] \\ + \mu_e \times \frac{\partial}{\partial x}\left[-n(x) \times \frac{\partial V(x)}{\partial x} + k_B T \times \frac{\partial n(x)}{\partial x}\right] \quad (18)$$

$$\frac{\partial p(x)}{\partial t} = k_{\text{dis}} \times [\xi(x) - \xi_0] - [n(x) - n_0] \times [p(x) - p_0] + \mu_h \times \frac{\partial}{\partial x}\left[p(x) \times \frac{\partial V(x)}{\partial x} + k_B T \times \frac{\partial p(x)}{\partial x}\right] \quad (19)$$

$$\frac{\partial \xi(x)}{\partial t} = \beta_{\text{corr}} k_{\text{Ex-CT}}(x) \times [\chi(x) - \chi_0] + k_{\text{rec}} \times [n(x) - n_0] \times [p(x) - p_0] - k_{\text{dis}} \times [\xi(x) - \xi_0] \\ - k_{\text{CT}} \times [\xi(x) - \xi_0] - k_{\text{CT-Ex}}(x) \times [\xi(x) - \xi_0] \quad (20)$$

$$\frac{\partial \chi(x)}{\partial t} = G_{\text{Ex}}(x) - \beta_{\text{corr}} k_{\text{Ex-CT}}(x) \times [\chi(x) - \chi_0] - k_{\text{Ex}} \times [\chi(x) - \chi_0] + k_{\text{CT-Ex}}(x) \times [\xi(x) - \xi_0] \quad (21)$$

$$J_n(x) = \mu_e \left[-n(x) F(x) - k_B T \frac{\partial n(x)}{\partial x}\right] \quad (22)$$

$$J_p(x) = \mu_h \left[p(x) F(x) - k_B T \frac{\partial p(x)}{\partial x}\right] \quad (23)$$

$$F(x) = \frac{\partial V(x)}{\partial x} \quad (24)$$

where $x$ is the position in the 1D simulation, $n$ is the density of electron, $p$ is density of holes, $\chi$ is density of CT states, $\xi$ is density of Ex states, $\mu_e$ is the electron mobility, $\mu_h$ is the hole mobility, $k_{rec}$ is the bimolecular recombination coefficient of free carriers to form CT states, $k_B$ is the Boltzmann constant, $T$ is the temperature, $k_{\text{Ex-CT}}$ is the transition rate from Ex states to CT states, $k_{\text{Ex}}$ is the relaxation rate of Ex states, $G_{\text{Ex}}$ is the generation rate of Ex states by photo excitation, $k_{\text{CT-Ex}}$ is the transition rate from CT states to Ex states, $k_{\text{CT}}$ is the relaxation rate of CT states, $k_{\text{dis}}$ is the dissociation rate of CT states, and $\beta_{\text{corr}}$ is the correction factor due to exciton diffusion[11,37,164]. In this simulation, Ex and CT states are assumed to be immobilized.

| Parameter | Value |
| --- | --- |
| $k_{\text{Ex}}$ (s$^{-1}$) | $1\times10^9$ to $1\times10^{11}$ |
| $k_{\text{rec}}$ (cm$^3$s$^{-1}$) | $2.5\times10^{-10}$ |
| $\mu_e$ (cm$^2$V$^{-1}$s$^{-1}$) | $5\times10^{-4}$ |
| $\mu_h$ (cm$^2$V$^{-1}$s$^{-1}$) | $5\times10^{-4}$ |
| $k_{\text{dis}}$ (s$^{-1}$) | $5\times10^{10}$ |
| $k_{\text{CT}}$ (s$^{-1}$) | $1\times10^9$ |
| $G_{\text{Ex}}$ (cm$^{-3}$s$^{-1}$) | from transfer matrix ($\sim 1\times10^{22}$-$4\times10^{22}$) |
| $\beta_{\text{corr}}$ | 0.1 |

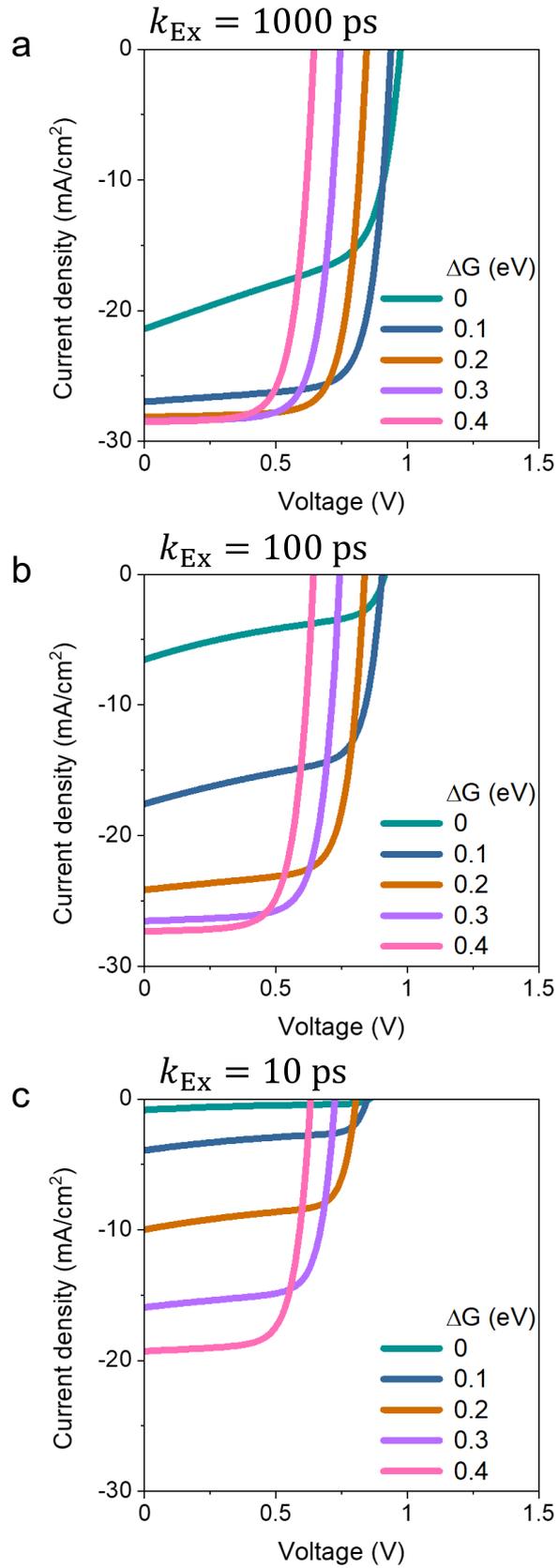

**Supplementary Figure 12 | The simulated JV curves.** Simulated JVs by using the Marcus theory and the Stark effect. to describe the field dependent charge transfer process from Ex to CT at different Ex decay rate. Here, $\Delta G = \Delta E_{\text{Ex-CT}} = E_{\text{Ex}} - E_{\text{CT}}$.

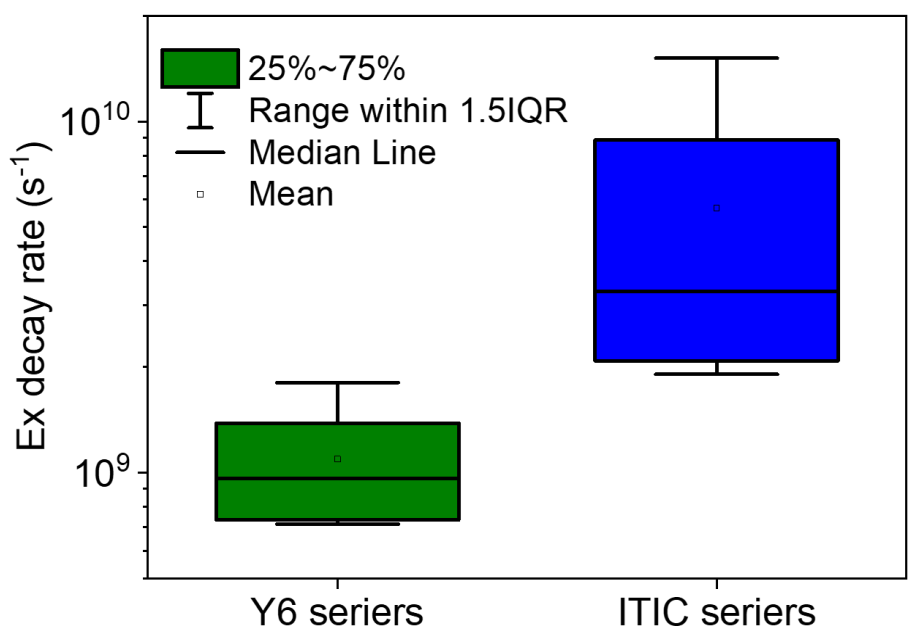

**Supplementary Figure 13 |** Exciton decay rate statistics of Y6-series (all with the thiazole core unit) and IT-series NFAs. IQR is short for the interquartile range.

| Thin film material | Lifetime (ns) | Source |
|---|---|---|
| Y6 | 1.016 | ref[11] |
|  | 1.4 | ref[60] |
|  | 1.38 | ref[61] |
|  | 0.8 | ref[62] |
|  | 1.36 | ref[63] |
|  | 0.723 | ref[64] |
| L8-BO | 0.672 | ref[65] |
|  | 1.06 | ref[66] |
| BTP-ec9 | 0.554 | ref[67] |
|  | 1.1 | ref[68] |
| IT-4F | 0.525 | ref[69] |
|  | 0.066 | ref[70] |
|  | 0.48 | ref[60] |
|  | 0.17 | ref[71] |
| ITIC | 0.305 | ref[11] |
|  | 0.40 | ref[60] |
|  | 0.113 | ref[70] |

**Supplementary Table 4 | The summary of exciton lifetime of NFA films.**

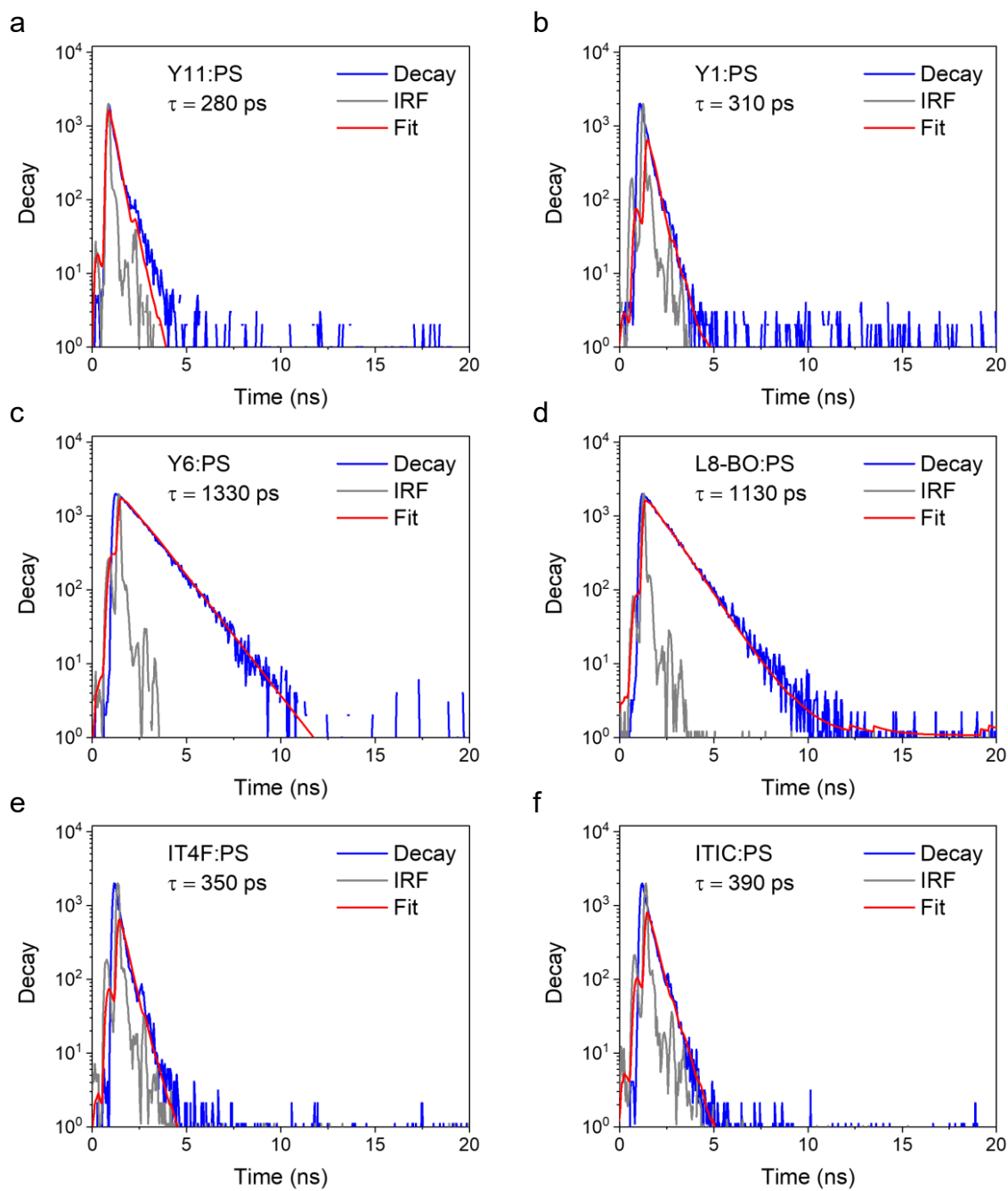

**Supplementary Figure 14 |** Time-resolved photoluminescence (TRPL) of NFA film samples with NFA dispersed in polystyrene (PS) and the weight ratio = 1:1. PS is an inert wide bandgap polymer used to reduce the difference in aggregation between pristine NFA films and the donor-acceptor blend films.